\documentclass[namedreferences]{solarphysics}

\usepackage[optionalrh]{spr-sola-addons} % For Solar Physics 
\usepackage{graphicx}        % For eps figures, newer & more powerfull
\usepackage{color}           % For color text: \color command
\usepackage{breakurl}        % For breaking URLs easily trough lines
\usepackage{booktabs}
\usepackage[utf8]{inputenc}

\chardef\us=`\_

\begin{document}

\begin{article}
\begin{opening}

\title{Iterative construction of the optimal sunspot number series}
\author[addressref={aff1,aff2},corref,email={svanda@asu.cas.cz}]{\inits{M.}\fnm{Michal}~\lnm{\v{S}vanda}\orcid{0000-0002-6345-1007}}
%\author[addressref={aff1,aff2},corref]{\inits{M.}\fnm{Michal}~\lnm{\v{S}vanda}\orcid{0000-0002-6345-1007}}
\author[addressref=aff1]{\inits{M.}\fnm{Martina}~\lnm{Pavelkov\'a}}
\author[addressref=aff3]{\inits{J.}\fnm{Ji\v{r}\'i}~\lnm{Dvo\v{r}\'ak}\orcid{0000-0003-3290-8518}}
\author[addressref=aff1]{\inits{B.}\fnm{Bo\v{z}ena}~\lnm{Solarov\'a}}
%\author{\inits{}\fnm{}~\lnm{}\orcid{}}
%   NOTE:  Just one corresponding author [corref]
%   \institute{$^{1}$ First affiliation
%                     email: \url{e.mail-a} email: \url{e.mail-b}\\ 
%              $^{2}$ Second affiliation
%                     email: \url{e.mail-c} \\
%             \textit{}
\address[id=aff1]{Astronomical Institute of the Czech Academy of Sciences, Fri\v{c}ova 298, CZ-25165 Ond\v{r}ejov, Czech Republic}
\address[id=aff2]{Astronomical Institute, Charles University, Faculty of Mathematics and Physics, V~Hole\v{s}ovi\v{c}k\'ach 2, CZ-18200 Prague, Czech Republic}
\address[id=aff3]{Department of Probability and Mathematical Statistics, Faculty of Mathematics and Physics, Charles University, Sokolovsk\'a 83, CZ-18675 Prague, Czech Republic}

\runningauthor{\v{S}vanda et al.}
\runningtitle{Iterative construction of the optimal sunspot number series}

\begin{abstract}
The relative number of sunspots represents the longest evidence describing the level of solar activity. As such, its use goes beyond solar physics, e.g. towards climate research. The construction of a single representative series is a delicate task which involves a combination of observation of many observers. We propose a new iterative algorithm that allows to construct a target series of relative sunspot number of a hypothetical stable observer by optimally combining series obtained by many observers. We show that our methodology provides us with results that are comparable with recent reconstructions of both sunspot number and group number. Furthermore, the methodology accounts for the possible non-solar changes of observers' time series such as gradually changing observing conditions or slow change in the observers vision. It also provides us with reconstruction uncertainties. We apply the methodology to a limited sample of observations by \v{C}ESLOPOL network and discuss its properties and limitations. 
\end{abstract}

\keywords{Solar Cycle, Observations; Sunspots, Statistics}
\end{opening}
%-------------------------------------------------

% \usepackage{amsmath}
% \usepackage{graphicx}
% \usepackage[colorlinks=true, allcolors=blue]{hyperref}

\section{Introduction}
``Counting of sunspots'', usually accompanied with the drawing of the photosphere, belongs to the most archaic methods used in nowadays's research. Yet, this longest-lasting ``measurement'' provided the mankind with important discoveries about our star. The discovery of 11-year \citep{1844AN.....21..233S} and other cycles \citep[see a review by][]{2010LRSP....7....1H} belongs to within the most widely known. Several ``by-products'' also need to be mentioned here. Let us mention e.g. the discovery of solar rotation itself (noted already by Galileo Galilei), learning that the Sun does not rotate like a solid body but reveals a differential rotation \citep{1630rour.book.....S,1860MNRAS..20..254C}, or the discovery of some of the empirical laws of solar activity, such as the Sp\"orer's law \citep{1858MNRAS..19....1C,1880POPot...2....1S}.

The counts of sunspots are the only direct records at our disposal to retrace the long-term changes in the solar cycle. The overall level of solar activity is closely related to the solar forcing acting on Earth's climate \citep[see e.g. reviews by][]{2013ARA&A..51..311S,2021RAA....21..131C}. Therefore, numbers of sunspots and their evolution in time consist a material that is used far beyond solar-physics applications. 

A systematic approach was founded by \cite{wolf1851sonnenflecken}, who started to record daily observations of the Sun at Z\"urich observatory, he later extended his records towards previous years (Wolf, 1861). It was known by that time that sunspots usually do not appear in the photosphere of the Sun isolated and form nests, the groups of sunspots. The discussion on whether it is the count of groups or the count of individual spots which is more important to assess the level of solar activity was worked around by Wolf, who introduced his ``relative number of sunspots'' $R$ by combining the counts of both quantities into one index. He weighted the number of groups ten times more as compared to the total number of sunspots, as he realised that it is easier for observers to identify sunspots groups. The weighting by a factor of 10 stemmed from the fact that according to his observations, there were \emph{on average} about 10 sunspots in one group. 

Wolf's goal was to have daily observations of the Sun, which was of course virtually impossible for a single observer at a single station. Hence Wolf already established a network of auxiliary observers, where he took their observations when his own were unavailable. In order to do so, he introduced a multiplicative coefficient $k$ that linked his observations to records of other observers so that the magnitudes of the relative numbers matched. This practice continued for years, where the procedure was such that there always was a primary observer and the observations of secondary or even tertiary observers were considered only when the primary observations were missing. The chain of primary observers is the origin of the backbone method \citep[][]{2016SoPh..291.2653S} of combining records of observers together into one composite, which today is used primarily when working with historical data.

In the modern era the abundance of the available observers is much larger, the composite may then be formed using completely different methods, usually based on some sort of averaging. For instance, the International Sunspot Number is derived from a network of stations worldwide using the following two-step algorithm \citep[see][]{2007AdSpR..40..919C,2019ApJ...886....7M}.

In the first step, the personal conversions coefficients $k$ are computed with respect to the records obtained at the Locarno pilot station. This is to assure the continuity, because Locarno observing station was used both in Z\"urich and Brussels era (the transition occurred in 1981). For each station, the observations on days when the daily $k$ coefficient deviates by more than $2\sigma$ from the monthly mean are considered outliers and rejected. This is repeated iteratively until the $k$ coefficients lie within the chosen statistical bounds from their monthly means. 

In the second step, the transformed $R$ value is computed for each station using the $k$ coefficient determined in the first step and for each day, the network average and its standard deviation is determined. The daily average is compared to the Locarno reference. Should the difference between Locarno pilot value and the network average exceed the standard network deviation, the Locarno reference is replaced by the network average and the $k$ coefficients are recomputed again. Finally, for each day the station values are confronted with the daily network average. Should the $k$-coefficient-transformed relative number differ more than $1\sigma$ from the network average, the exceeding station value is rejected and the daily average is computed again. This down-selection of observations is repeated again until a stable solution is found. The network daily average is then considered as a final daily relative number $R_i$. 

Additional sunspot numbers exist, differing slightly by methodology of their calculation. For instance, the Boulder Sunspot Number is derived from the daily Solar Region Summary produced by the US Air Force and National Oceanic and Atmospheric Administration and utilised sunspot drawings obtained from the Solar Optical Observing Network. The Boulder Sunspot Number was typically about 60\% larger than the International Sunspot Number until a large revision of the Internation Sunspot Number in 2015. In the US, another sunspot number representative \citep{1985JAVSO..14...28T} is provided by the American Association of Variable Star Observers (AAVSO).  

All these data sets differ by details and sometimes in magnitude, they can be considered as independent description or representation of the solar reality. In some of these relative numbers the issues were discovered. For instance, systematically increasing deviations were observed between the International Sunspot Number and the AAVSO number, which were accounted to the flaw in the AAVSO methodology \citep{1997JAVSO..26...40S,1997JAVSO..26...50F}.

Even the International Sunspot Number was largely revised recently when several issues were discovered and corrected in the historical records \citep{2014SSRv..186...35C} and even some non-solar trends in the relative numbers in recent decades \citep{2016SoPh..291.2629C}. This led to a new revised sunspot number \citep{2016SoPh..291.2629C} issued by the World Data Center Sunspot Index and Long-term Solar Observations (WDC-SILSO\footnote{\url{https://www.sidc.be/silso/}}), which replaced the International Sunspot Number. For the modern-era the independent sunspot numbers now agree quite well. The corrections to the historical records performed recently on International Sunspot Number by \cite{2014SSRv..186...35C} were not accepted by the entire community \citep[see e.g. section 3.4 of][]{2022SoPh..297....8V}. The Sunspot Workshops community\footnote{\url{https://ssnworkshop.fandom.com/wiki/Home}} decided that since new data and new techniques are retrieved and developed regularly, the sunspot number and other series should be included in a modern versioning system, like all scientific series.

Various methods of combining many independent observations with unknown uncertainties together to form one composite series suffer from various drawbacks. For instance, the backbone method is prone to jumps or drifts of the primary observer \citep[see e.g.][]{2016ApJ...824...54L}. Weighted averages may also lead to unobserved drifts. To avoid possible issues, it has been a common practice not to consider all available observations, but to do reject some observations by e.g. judging the stability of such observer (see e.g. the description of the WDC-SILSO algorithm above). These criteria often follow some rule of thumb instead of a rigorous assessment, such as e.g. a pre-selection based on the total number of observations by a particular observer by AAVSO network \citep[][]{1985JAVSO..14...28T}.

We propose a methodology of combining time series of several observers together when using as many observations as possible, not excluding any of them explicitly from the beginning and not rejecting any observations during the process. The proposed methodology is developed to be suitable for the networks of several observers, where the number of available daily observations is too small to apply the WDC-SILSO method with outlier rejection. The methodology targets the construction of the series of the hypothetical observer that optimally combines the recordings of the real observers. The methodology is based on an weighted-average approach, where the optimal combination is obtained by using an iterative procedure. This approach allows us to suppress the effects of the existing long-term drifts in individual series. This paper is a first step in the development of a method that will be further tested when a more rich set of observations will be available.

\section{Observations}

\subsection{ČESLOPOL}
Solar observations have a very long tradition in the countries appearing in the area of the nowadays Czech Republic \citep{2021Apis..127....6P}. Already from the beginning of the 17th century, drawings of the Sun were recorded by Johann Zimmermann, who was the first known systematic solar observer in the Czech countries. As of beginning of 19th century, several long-term systematic observers appeared, such as Gregor Mendel or Artur Kraus. With the foundation of the Czech Astronomical Society in 1917 solar observations became a common task among amateur observers within the activities of its Solar Section. In 1964 sunpots drawings were officially politically declared as a task of a national importance and in 1978 a FOTOSFEREX project was initiated \citep{1978Kozmo...9...85K,1978Rise...59...95K}. 

The goal of the project was to have daily sunspot drawings available, sunspot drawings became the core for the solar-activity predictions issued by the Astronomical Institute of the Czechoslovak Academy of Sciences. One has to bear in mind that those days the access to the world data was strongly limited due to the communist regime and the only solution was a foundation of a local network. FOTOSFEREX project relied heavily on local observations in Ond\v{r}ejov, only in the case of unfavourable observing conditions drawings from auxiliary stations (which included the amateurs) were considered. These drawings were sent by telex service to Ond\v{r}ejov. 

With the advent of synoptic programs such as SoHO \citep{1995SoPh..162....1D} or GONG \citep{1996Sci...272.1284H} the interest in using the solar observations obtained by non-professionals decreased significantly and project FOTOSFEREX was shut down in 1998. To promote solar observations in recent years, project ČESLOPOL was initiated.

ČESLOPOL stands for ČESko-SLOvensko-POLská (Czech-Slovak-Polish) network of solar observers working for the Solar Section of the Czech Astronomical Society. At the moment, it includes 24 active observing stations. The network collaborates closely with the Solar patrol of the Astronomical Institute of the Czech Republic (ASU). Not only it gathers current observations of the Sun, however, it targets to include the complete archive of FOTOSFEREX and the archive of the national task and also to digitize and evaluate historical drawings. The aim of the project ČESLOPOL is to integrate all the sunspot drawings obtained in the historical Czech countries into one large database. 

\subsection{Data}

Observers included in the ČESLOPOL network observe and draw the Sun in the usual way, which is well known among amateur and professional astronomers, usually using the projection technique. The drawings are evaluated also in a standard way, their digitisation is done by a custom software {\sc Slunce}\footnote{The program is available for download from \url{https://www.asu.cas.cz/~sunwatch/cs/stranka/ke-stazeni}, however its interface is in the Czech language.}. This is a GUI-based software with a convenient user interface that allows to simply input semi-evaluated drawing of the photosphere and computes and stores the data in a digitised form. It also allows to export the digitised observations in a CSV (comma separated values) format to allow a communication with the outer world. 

The archive of digitised observations grows every day. Already it is large enough to think about a calculation of an optimal representative relative sunspot number in an automatic way. For the purpose of this study we had the following subsets at disposal:

\begin{itemize}
    \item A subset of observations by Ladislav Schmied covering years 1986--1993 from his private observatory in Kun\v{z}ak. His total personal archive covers years 1947--2012, however, most of these observations were not digitised yet. 
    \item Observations from Solar Patrol of the Astronomical Institute of the Czech Academy of Sciences from Ond\v{r}ejov observatory, covering years 1986--2020. Over the years, two main observers were active there for a long term, Franti\v{s}ek Zloch and Tom\'a\v{s} Van\v{e}k, complemented by several other observers active for shorter times, including the authors of the study. We only add that at the moment, observations from years 1949--2022 were digitised and are in principle available. 
    \item Observations from Pre\v{s}ov observatory, Slovakia, 1984--2009. Again, time span is larger, not all the data were properly digitized. 
    \item Recordings form the Observatory Prost\v{e}jov in years 1991--2019.
    \item Scarce observations from Observatory Uhersk\'y Brod distributed over years 1988--2004. 
\end{itemize}

\begin{figure}
    \centering
    \includegraphics[width=\textwidth]{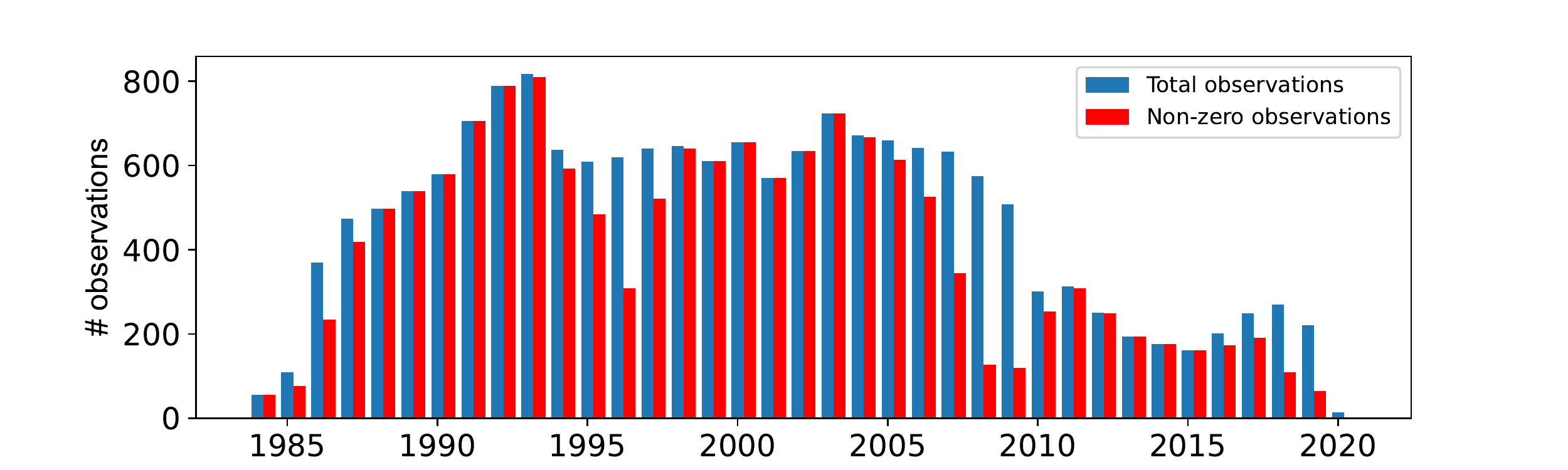}
    \caption{Annual counts of individual observations available in the tested set (blue) together with the count of non-zero observations (where $R>0$).}
    \label{fig:count_observations}
\end{figure}

In total, the data archive used in this study contains 16302 individual digitised sunspot drawings performed by 64 individual observers operating in 5 stations. Yearly counts of processed drawings including the count of non-zero observations (that is where $R>0$) are given in Fig.~\ref{fig:count_observations}. The plot demonstrates  that the count of observations per year varies, it also may depict a long-term trend. 

\begin{figure}
    \centering
    \includegraphics[width=\textwidth]{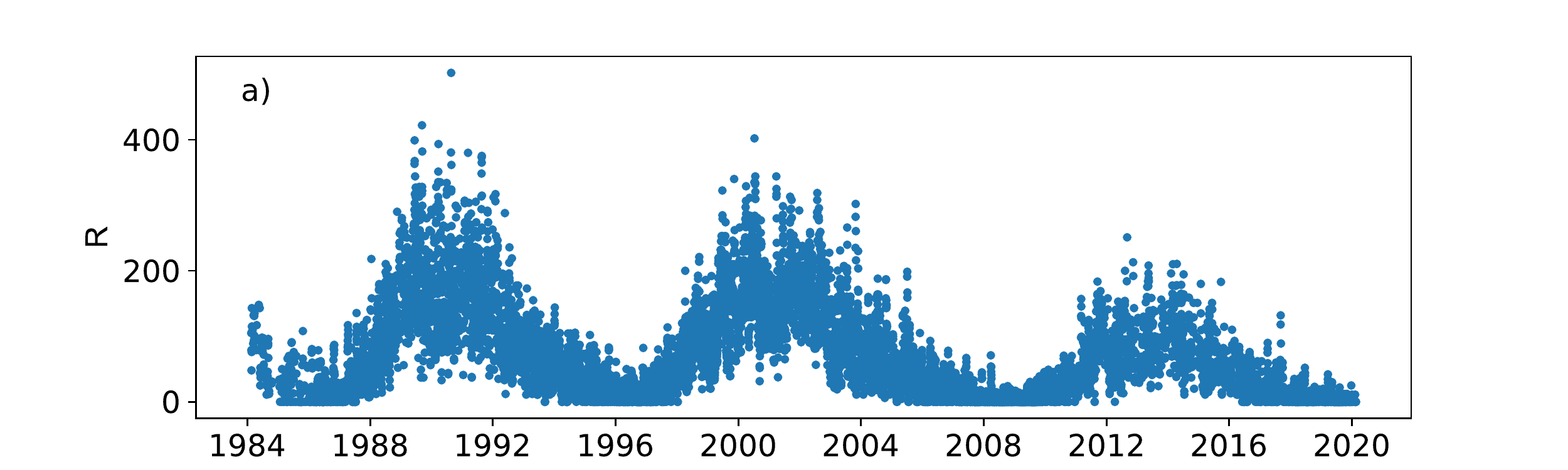}
    \includegraphics[width=\textwidth]{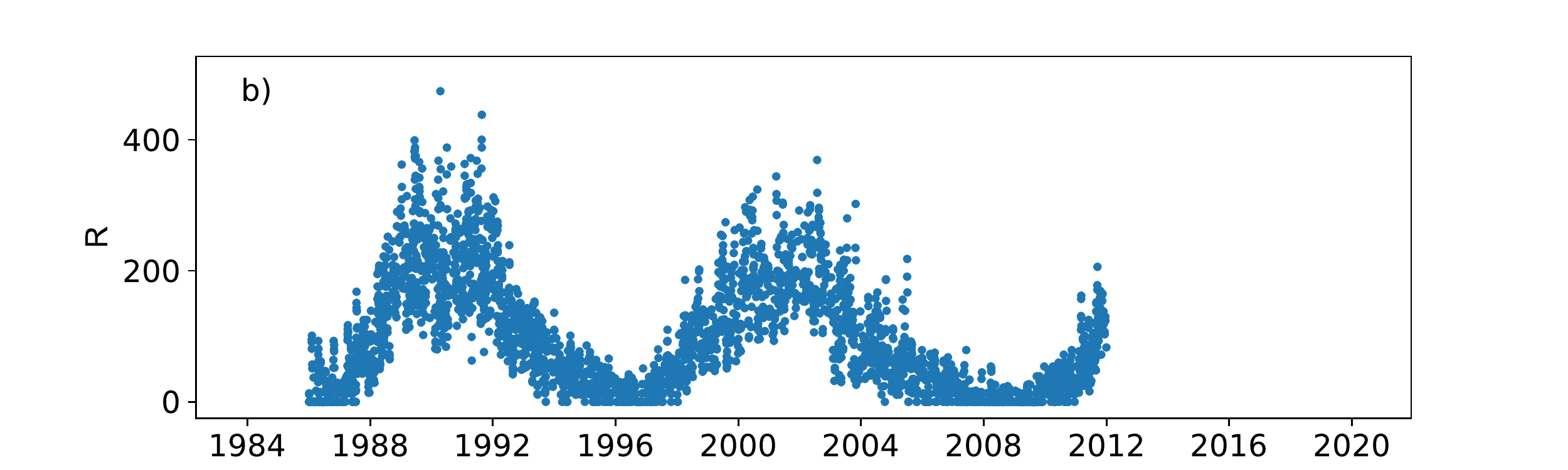}
    \includegraphics[width=\textwidth]{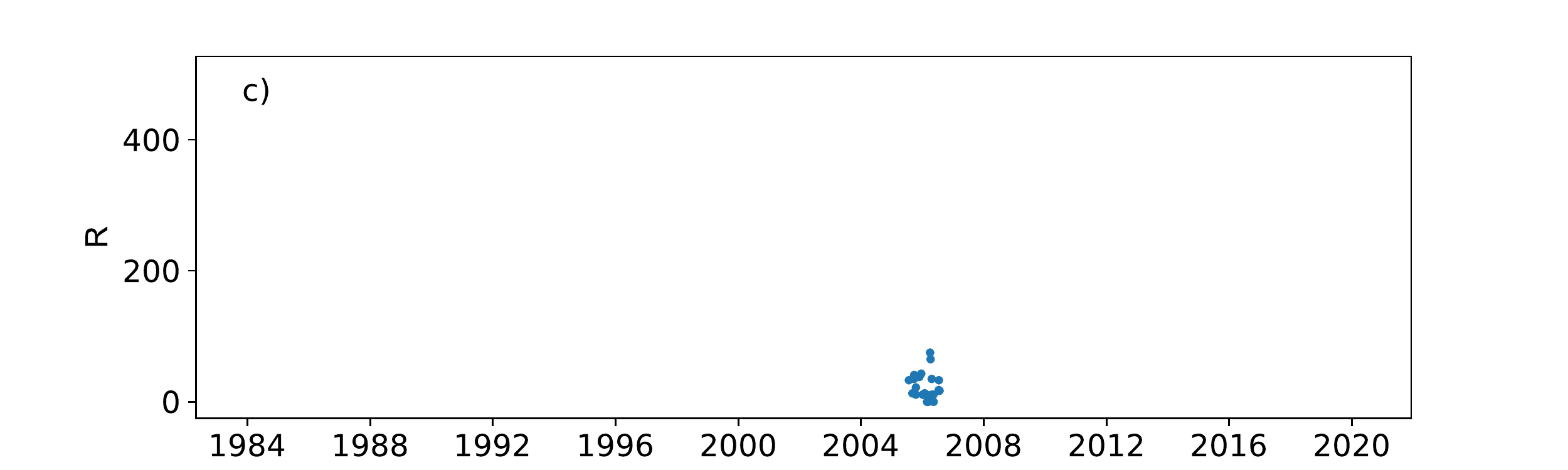}
    \caption{Demonstration of the observations used for this study. a) -- A simple daily mean of all used observations. b) -- An example history of an excellent long-term observer František Zloch. c) -- An example of an episodic observer (the corresponding author of this study served at the Solar patrol during his Ph.D. only weekend services). The scales and ranges of all panels are intentionally identical. }
    \label{fig:data_examples}
\end{figure}

\subsection{Data properties}
Among other, three values from the records are most important for our study: the number of recorded sunspot groups $g$, the total number of recorded individual sunspots (umbrae) $f$ and the relative sunspot number $R$, which depends on the former two. Example plots of $R$ from the archive are given in Fig.~\ref{fig:data_examples}, where we plot first a simple average (average of $R$s of individual observations obtained on the same day) of our archive and also two examples of observers working for Solar patrol ASU. The figure shows that the input data are not continuous and that the coverage by various observers may differ significantly. However, each of these observers may cover some important part of the final series, so we do not find it wise to exclude some of them from the beginning. We only point out that case c) in Fig.~\ref{fig:data_examples} would very likely be excluded from algorithms constructing the WDC-SILSO or AAVSO sunspot numbers due to the short coverage. The inclusion of even such short series allows us to work with individual observers rather than with observing stations, which is the practice e.g. by WDC-SILSO, considering that even using the same equipment the personal biases may be different for different human beings. 

One of the key factors influencing the ``closeness'' of the drawing to the solar reality are the observing conditions. In ČESLOPOL network a five-grade scale of quality of observing conditions $Q$ is established. It is not a rigorous scale, it describes subjectively the level of details that are visible during the observation, see Table~\ref{tab:Q_table}. Already from the definitions of individual $Q$-values it is clear that the number of recorded sunspots and possibly groups (especially when discussing the single-pore groups of McIntosh classification Axx) strongly depends on the value of $Q$. The smaller the value, the lesser number of sunspots will likely be captured and drawn. 

\begin{table}[]
    \centering
    \begin{tabular}{c|l}
    \toprule
    $Q$ & Definition\\
    \midrule
    1 & only large spots are visible, granulation is not visible; poor observing conditions\\
    2 & also smaller spots are visible, granulation visible only occasionally\\
    3 & smaller spots are well visible, granulation visible; average observing conditions \\
    4 & granulation and pores visible well\\
    5 & all the details are observable; exceptional observing conditions\\
    \bottomrule
\end{tabular}
    \caption{Subjective scale of observation quality $Q$}
    \label{tab:Q_table}
\end{table}

\begin{figure}
    \centering
    \includegraphics[width=\textwidth]{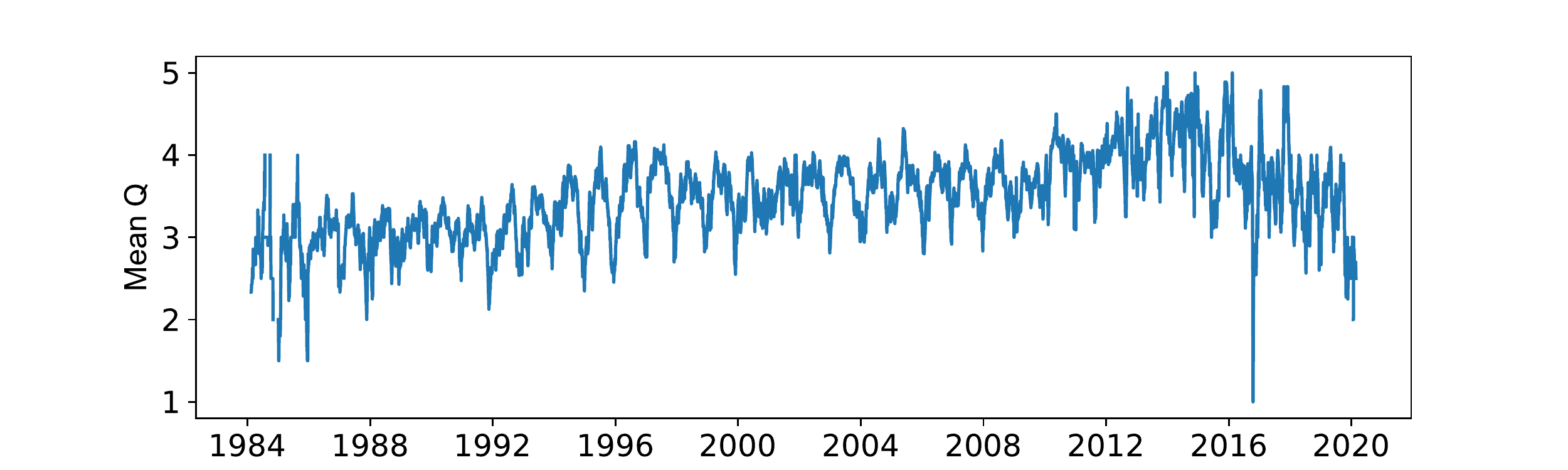}
    \caption{The values of $Q$ smoothed by the running average over 13 days. }
    \label{fig:meanQ}
\end{figure}

First, we studied the long-term evolution of the reported observing conditions by taking the mean of all observations available for a particular day. In Fig.~\ref{fig:meanQ} one can see clear oscillations with a period of one year, which record the seasonally changing observational conditions. On average, the observing conditions are better in summer, where usually the drawing was recorded early in the morning under decent observing conditions. The conditions gradually worsen towards winter, when in the Czech Republic the weather is usually not favourable for observations. Cloudy skies are often a reality during the days at the end and beginning of the calendar year. The pattern of annual oscillations is quite regular until say year 2009, in the years onwards the behaviour seems more chaotic. We believe that this gradual change has to do with a decreasing number of available observations after 2010, which is evident from Fig.~\ref{fig:count_observations}. The \emph{average} $Q$ suffers from the small-number statistics here. 

In Fig.~\ref{fig:meanQ} one also notices a gradual growth of average $Q$ in years 1984--2014 followed by a decrease onwards. In years 1984--1994 the average $Q$ oscillates around the value of 3, which stands for \emph{average} observing conditions, hence the scale given in Table~\ref{tab:Q_table} corresponds to its definition. Between years 1998 and 2010 the mean $Q$ is about 3.5 and in years 2012--2015 it even reaches the mean value of almost 4. After 2018 the mean $Q$ is about 3.5 again. It is not clear as to why the observing conditions should slowly improve over the long period and then start to worsen again. 

\begin{figure}
    \centering
    \includegraphics[width=\textwidth]{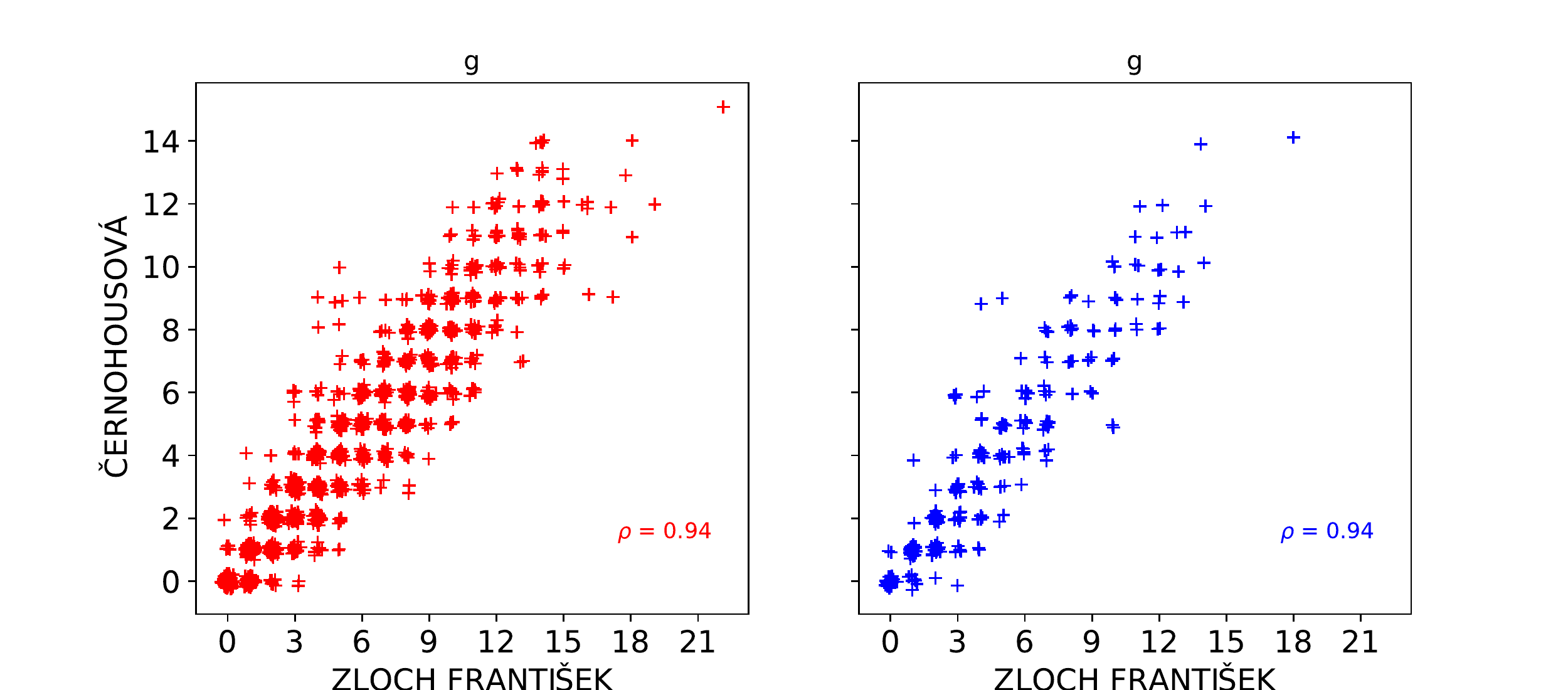}
    \includegraphics[width=\textwidth]{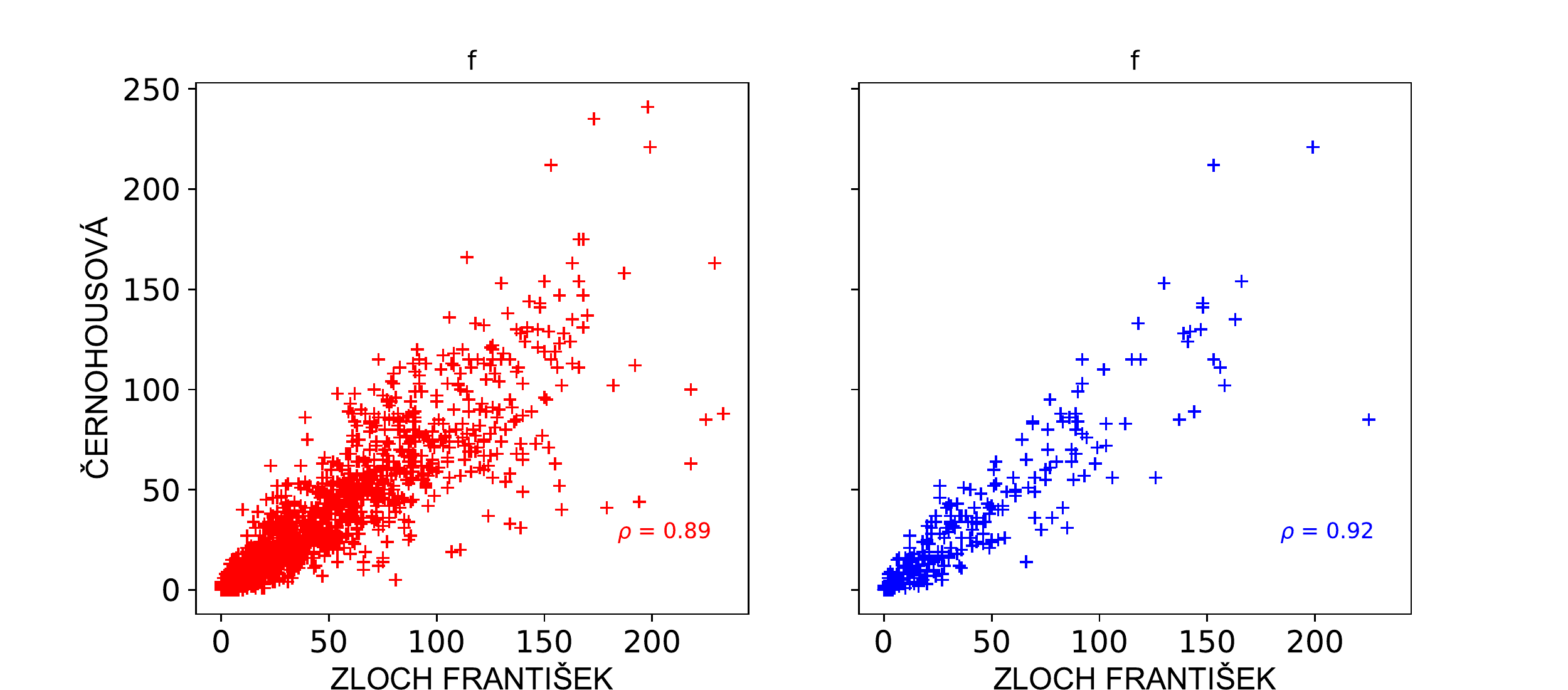}
    \includegraphics[width=\textwidth]{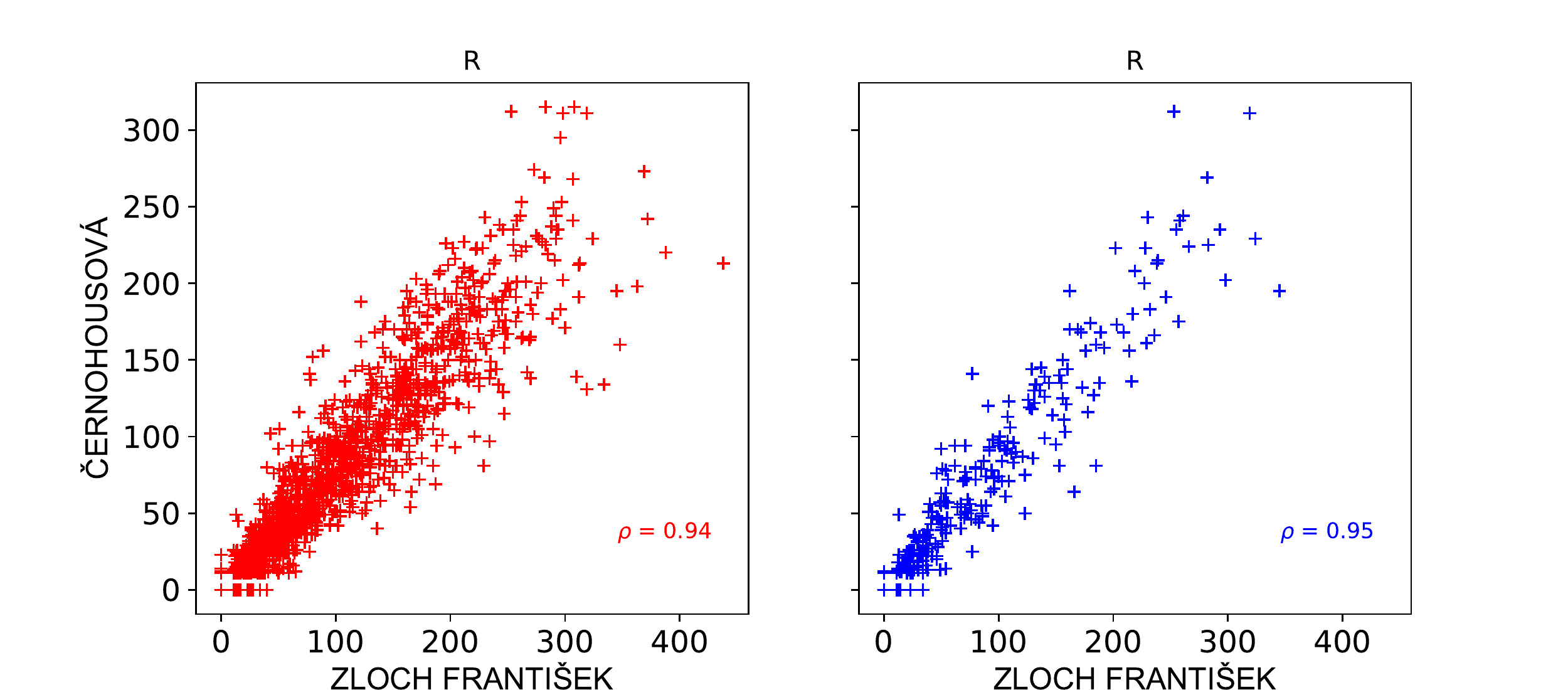}
    \caption{Example of direct comparison of two observers. Each cross corresponds to the value obtained by the two observers at the same day (red) and at the same day and with the same declared conditions (blue). The corresponding correlation coefficients are given in the plot in the corresponding colours. Counts of groups, spots, and the reported relative number are plotted in individual panels. In the displayed case, the studied overlap period was more than 20 years and included 286 ``blue'' and 1561 ``red'' days to be compared. Note that in the plots in the upper row we added a small random jitter to the value to avoid point overlapping. }
    \label{fig:Q_day}
\end{figure}

If these trends in $Q$ are real, they must introduce a gradual change in the recorded values of $g$, $f$, and consequently $R$. We keep the usual definition of the relative number $R$ as obtained from the observations,
\begin{equation}
    R=10g+f,\label{eq:R}
\end{equation}
where $g$ is the count of sunspot groups and $f$ is the total number of sunspots. When combining the dataset from various observers a personal coefficient $k$ is introduced, which is used traditionally to scale personal relative number $R$ to the ``representative'' one $R_{\rm i}$:
\begin{equation}
    R_{\rm i}=kR=k(10g+f). \label{eq:Ri_R}
\end{equation}
The coefficient $k$ includes personal/instrumental biases of the observer. Not only it scales the observations of an observer to the reference data series, however, in general it also scales one observer to another (the value naturally depends on the selection of the two series to be compared). Hence the observations of two observers should be bound by a single scaling coefficient. We test this idea in two ways: first we chose the observations performed by two observers at the same days and compare not only the derived relative numbers $R$, but also the count of sunspot groups $g$ and the total count of sunspots $f$. Then we do the same with a more limited sample for the same two observers, for observations obtained at the same day and with the same declared observing conditions. Example plots for one pair of tested observers it given in Fig.~\ref{fig:Q_day}.

We find that the idea of the coefficient scaling is valid fairly well, however the scaling coefficient seems to differ for number of sunspot groups and for the total number of sunspots. In some cases we found that these coefficients may differ even three times. We speculate that this may have to do with the effective resolution of the used telescope, when it may be easier to identify the groups, however to properly resolve individual spots within the groups may be more difficult. 

The scatter plots show a wider distribution along the hypothetical linear fit in the case when different observing conditions were allowed for cross-comparison than when the same observing conditions were requested. Also the correlation coefficient is usually smaller in the former case. 

This simple test shows that the observing conditions play a role in deriving the personal scaling coefficient. This justifies the need for transformation of the obtained observations to optimal ($Q=5$) observing conditions, even though the effect is small. To achieve the goal we propose to transform $g$ and $f$ to standard conditions by simple linear transforms,
\begin{eqnarray}
g_{\rm red}&=&g[1-c_g (5-Q)],\label{eq:gred}\\ 
f_{\rm red}&=&f[1-c_f (5-Q)].\label{eq:fred}
\end{eqnarray}
The values of linear-fit coefficients $c_g$ and $c_f$ were obtained for each observer independently. In order to have some meaningful estimate we considered only 8 observers from our set with the largest number of observations. For each of these observers we considered linear relations similar to (\ref{eq:gred}) and (\ref{eq:fred}), where $g$ and $f$ values represented those of the observer and $g_{\rm red}$ and $f_{\rm red}$ were taken from the international sunspot number SILSO v2.0 for the same days\footnote{In principle, the coefficients might be derived with respect to \emph{any} reasonable description of $g$ and $f$. Another option than SILSO sunspot number would be for instance a simple mean as plotted in panel a) in Fig.~\ref{fig:data_examples}. Another option would be to perform the process iteratively: 1. set both $c_g$ and $c_f$ to zero, 2. perform the optimisation run as described by our methodology, use the resulting series as the reference, 3. determine $c_g$ and $c_f$ with respect to this new reference and 4. run a final optimisation.}. We realise that the definition of $Q$ is somewhat fuzzy, hence we assumed an uncertainty of 0.5 for each $Q$ value in the least-squares linear fit. Furthermore, only non-zeros values of $g$ and $f$ were considered in the fit.

The values of $c_g$ and $c_f$ obtained for 8 observers show a spread, however they are all negative. This fact confirms the assumption that the reduction of observations to \emph{better} observing conditions must increase the $g$ and $f$ values, as if the observer would see more spots and more groups under the better conditions. The representative values of $c_g=-0.0348$ and $c_f=-0.0780$ with standard deviations of $0.0046$ and $0.0190$ respectively were obtained from the weighted average of the fits for the 8 observers, the weights reflected the number of observations considered in the individual fits. The reduction to optimal observing conditions is one of the novel points of our methodology. 

The subjective or somewhat objective judgement or description of the observing conditions may be different for other sunspot-drawing collection projects. We believe that our approach of transforming the relative numbers to fixed (e.g. optimal) observing conditions may be modified accordingly. 

Now that we have the reduction coefficients for (\ref{eq:gred}) and (\ref{eq:fred}), we are able to transform observations of all observers to optimal observing conditions and invoke them in the construction of the optimal series of relative sunspot number. Finally we note that the values of $g_{\rm red}$ and $f_{\rm red}$ are kept with a floating precision for further use and not rounded to integers as one could expect from their natural interpretation. 

\section{Methodology}
In this study we propose an iterative method of calculation of the representative relative sunspot number from a set of observations delivered by different observers. Our methodology does not prefer any of these observers and does not exclude any of them at the same time. It uses all observations and constructs the representative series in an optimal way. 

The premise is that all the observers observe the same Sun. However, due to their personal/instrumental biases and local possibly varying observing conditions their recordings may differ not only from the real Sun, but also when records of different observers are mutually compared. The goal is to find the optimal estimate of the real situation on the Sun which mimics the best the recordings of individual observers and at the same time estimate the level of their personal/instrumental biases and their changes over time. In fact, we construct a time series of a \emph{hypothetical observer} (who corresponds to the primary observer in the backbone method) by taking all available observations into account not preferring any of the real observers. The aim is to remove the dependence on the reference observer (which is the core of the backbone method). This is important because the personal/instrumental bias of the backbone may also change in time. 

\subsection{Conversion coefficients}
\begin{table}[]
    \centering
    \begin{tabular}{r|c}
         \toprule
         Way & $k_g$ \\
         \midrule
         Schmied Ladislav $\rightarrow$ Ivan&  0.375\\
         Schmied Ladislav $\rightarrow$ Vaněk Tomáš $\rightarrow$ Ivan & 0.296\\
         Schmied Ladislav $\rightarrow$ Zloch František $\rightarrow$ Ivan & 0.444\\
         Schmied Ladislav $\rightarrow$ Černohousová $\rightarrow$ Ivan & 0.501\\
         average (Schmied Ladislav $\rightarrow$ \emph{1 observer} $\rightarrow$ Ivan) & 0.416\\
         average (Schmied Ladislav $\rightarrow$ \emph{2 observers} $\rightarrow$ Ivan) & 0.449\\
         \bottomrule

         \toprule
         Way & $k_f$ \\
         \midrule
         Schmied Ladislav $\rightarrow$ Ivan&  0.800\\
         Schmied Ladislav $\rightarrow$ Vaněk Tomáš $\rightarrow$ Ivan & 0.600\\
         Schmied Ladislav $\rightarrow$ Zloch František $\rightarrow$ Ivan & 0.833\\
         Schmied Ladislav $\rightarrow$ Černohousová $\rightarrow$ Ivan & 1.000\\
         average (Schmied Ladislav $\rightarrow$ \emph{1 observer} $\rightarrow$ Ivan) & 0.811\\
         average (Schmied Ladislav $\rightarrow$ \emph{2 observers} $\rightarrow$ Ivan) & 0.849\\
         \bottomrule
         
    \end{tabular}
    \caption{Example of testing of the recalculation of count of groups $g$ and count of spots $f$ by recalculation from data series by other observers.}
    \label{tab:testing_k_idea}
\end{table}

\begin{table}[]
    \centering
    \begin{tabular}{r|c}
         \toprule
         Way & $k_g$ \\
         \midrule
         Vaněk Tomáš $\rightarrow$ Zloch František&  N/A\\
         Vaněk Tomáš $\rightarrow$ Schmied Ladislav $\rightarrow$ Zloch František & 1.389\\
         Vaněk Tomáš $\rightarrow$ Černohousová $\rightarrow$ Zloch František & 1.143\\
         Vaněk Tomáš $\rightarrow$ Ivan $\rightarrow$ Zloch František & 1.000\\
         average (Vaněk Tomáš $\rightarrow$ \emph{1 observer} $\rightarrow$ Zloch František) & 1.177\\
         average (Vaněk Tomáš $\rightarrow$ \emph{2 observers} $\rightarrow$ Zloch František) & 1.189\\
         \bottomrule

         \toprule
         Way & $k_f$ \\
         \midrule
         Vaněk Tomáš $\rightarrow$ Zloch František&  N/A\\
         Vaněk Tomáš $\rightarrow$ Schmied Ladislav $\rightarrow$ Zloch František & 1.511\\
         Vaněk Tomáš $\rightarrow$ Černohousová $\rightarrow$ Zloch František & 1.080\\
         Vaněk Tomáš $\rightarrow$ Ivan $\rightarrow$ Zloch František & 1.005\\
         average (Vaněk Tomáš $\rightarrow$ \emph{1 observer} $\rightarrow$ Zloch František) & 1.199\\
         average (Vaněk Tomáš $\rightarrow$ \emph{2 observers} $\rightarrow$ Zloch František) & 1.231\\
         \bottomrule
         
    \end{tabular}
    \caption{Similar to Tab.~\ref{tab:testing_k_idea} only for a pair of observers who did not have any common observing day because they alternated at Solar patrol in Ond\v{r}ejov.}
    \label{tab:testing_k_idea1}
\end{table}
The personal scaling coefficients are in fact a key in success of combination of series obtained by many observers into one representative series. The coefficients $k$ may change slowly in time. We realised that the two quantities $g$ and $f$ are way too different in nature that one scaling coefficient may not be enough to cover all possible observing issues. 
Instead, we use two personal coefficient $k_g$ and $k_f$ that take into account personal/instrumental biases in the conversion of the observers' records to the representative relative number $R_0$:
\begin{equation}
    R_0 = 10 k_g g + k_f f. \label{eq:R0_two_coefficients}
\end{equation}
Here again, both $k_g$ and $k_f$ are allowed to change in time. 

We would like to combine all sets obtained by individual observers together into the composite series (which we term the \emph{target series} hereafter) to robustly describe the real evolution of sunspot count on the Sun. The idea of the personal coefficients may easily be tested on a set of observers which have at least some overlapping days. This procedure requires a somewhat robust calculation of the conversion coefficients. In the case of more observers in the sample it should in principle be possible to compute the conversion coefficient of observers 1 and 2 by using an intermediate observer 3, when coefficients between observer 1 and 3 and observer 3 and 2 are known. This procedure is also termed ``daisy-chaining'' by \cite{2016ApJ...824...54L}. 

Our tests showed that the idea holds only approximately. We used the whole digitised archive for a selection of long-term observers to test the idea. Table~\ref{tab:testing_k_idea} shows a typical example, where the conversion coefficient between observer Ladislav Schmied and observer Ivan ($k_g=0.375$, $k_f=0.800$) are approximated satisfactorily when using one observer in the middle ($k_g=0.415$, $k_f=0.811$ on average) or even involving two-step observer chain in the middle ($k_g=0.449$, $k_f=0.849$). Some other tests between other observers turned out better with a better agreement, some worse. The tests usually showed that the values of $k_g$ and $k_f$ may be significantly different, thus justifying our use of these separate coefficients. 

We have to stress out that in this test we computed the values of the conversion coefficients for all the overlapping observations, hence we did not allow for the change of the conversion coefficients in time. Also the number of mutual pairs of observations suitable for the computation of the conversion coefficients varied among observers, yet, we did not introduce any weighting. Given the simplifications we consider the results of these tests not excellent but satisfactory. Finally, we would like to point out that the methodology in principle allows to determine the (virtual) conversion coefficients of two observers who did not have any overlapping observations (see e.g. Table~\ref{tab:testing_k_idea1}). In our database this was the case for instance of observers alternating at Solar patrol at Ond\v{r}ejov observatory, where only one drawing per day was considered for the archive and hence these observers never made drawings on the same day. Yet, the virtual conversion coefficients when using one observer in the middle and two observers in the middle provide with a very reasonable match. 

\begin{figure}
    \centering
    \includegraphics[width=0.6\textwidth]{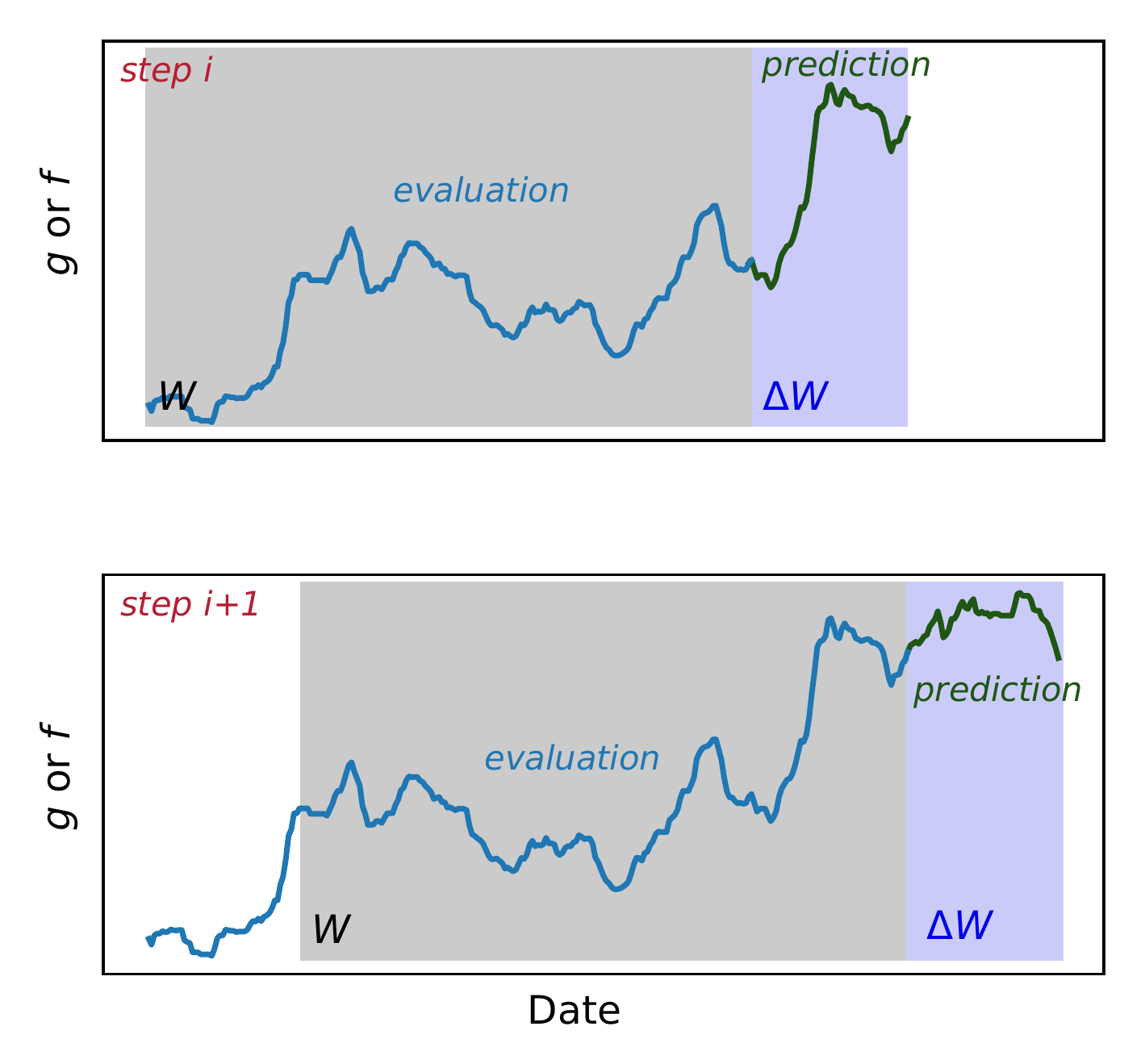}
    \caption{Schematic cartoon illustrating two consecutive steps of the method. Within the window $W$ the coefficients of all available observers (within that window) and the reference series are evaluated. In the following window $\Delta W$ these coefficients are use to predict what would the reference series be taking the observations of other observers into account.}
    \label{fig:method}
\end{figure}

\subsection{The algorithm}
Most of the relative sunspot number composites rely on a pilot station to ensure the continuity. The pilot station then holds a special status in the observers network. We would like to drop this need of the pilot station, because e.g. in the network of amateur observers, it is difficult or even undesirable to define a pilot station. We would also like to split observing teams at stations to set of individuals. Unfortunately, there still is a need to calibrate one observer to another by using the personal $k$ coefficients. Unlike in the procedure used in WDC-SILSO, we would like to separate the calculation of the personal conversion coefficients from the construction of the network composite, at least so that these two different steps do not use the same set of observations.
At the same time we target the usage of our code to networks with a few observers, maybe few dozens of them, definitely not as many as are included in the WDC-SILSO network. 

To compute a target sunspot number series by taking into account observations from a set of observers we propose an iterative methodology. It is schematically displayed in Fig.~\ref{fig:method}. The methodology is in principle divided in two alternating major steps: the period of evaluation (of the personal coefficients) and a subsequent period of prediction (of expected number of groups and sunspots). In both of the periods we work with the target series $g_{\rm 0}$ and $f_{\rm 0}$, from which the target relative number $R_0$ is computed by using (\ref{eq:R}). The target series $g_0$ and $f_0$ is built piece-by-piece during the program run. 

At start the target series does not exist yet, so for the evaluation of the first set of personal coefficients an initial series is used. The sensitivity of the procedure to selection of the initial series will be discussed in detail later in the paper (Section~\ref{sect:initial}). At this moment a natural choice would be some series, which is sufficiently ``solar-like''. The use of the observations of a long-term observer, who is expected to be somewhat stable, should be sufficient. During the program run, the segments of the target series provisionally filled by the initial series are replaced by the properly computed values of the target. 

In the period of evaluation, which is performed in a time window having a length of $W$, we follow the logic outlined by (\ref{eq:R0_two_coefficients}). For each observer $\alpha$ who provided observations in the evaluation period, we compute her/his personal coefficients $k^\alpha_g$ and $k^\alpha_f$, which optimally scale his/her values to the target series. Each and every observation $j$ that yields a pair of values $g^\alpha_j$ and $f^\alpha_j$ is first reduced to the optimal observing conditions by (\ref{eq:gred}) and (\ref{eq:fred}). To improve the stability of the evaluation period, we consider only those observations where $g^\alpha_i>1$ and $f^\alpha_i>5$. 

Then for each observation we compute the instant conversion coefficients
\begin{equation}
    k^\alpha_{g,j}=g^\alpha_{j,\rm red}/g_0\quad {\rm and}\quad k^\alpha_{f,j}=f^\alpha_{j,\rm red}/f_0. \label{eq:k_W}
\end{equation}
The representative coefficients for the evaluation period having length $W$ are then computed as an arithmetic average over all determined instant coefficients, simply as: 
\begin{equation}
    k^\alpha_g=\frac{1}{M^\alpha} \sum\limits_{j \in W} k^\alpha_{g,j} \quad {\rm and}\quad k^\alpha_f=\frac{1}{M^\alpha} \sum\limits_{j \in W} k^\alpha_{f,j},
\end{equation}
where $M^\alpha$ is a total number of observations of a given observer within the evaluation window fulfilling the thresholding condition given above. For each observer we thus obtain representative (within the window of length $W$) scaling coefficients and we keep also number $M^\alpha$ which will serve as a measure of the quality of determination of these coefficients. 

Then we move forward to the prediction period. This occurs in a window of length $\Delta W$, which immediately follows the evaluation period in time. In this period we ``predict'' the target series from the observations of other observers by using their personal conversion coefficients, which were determined during the previous period of evaluation. 

In the prediction window having length of $\Delta W$ we progress on a day-by-day basis. For each day $D$ we identify a subset of observers with personal conversion coefficients from the previous evaluation period that provide observations for this particular day. By using an observer $\beta$\footnote{Note that a set of identifiers $\beta$ forms a subset of the set of identifiers $\alpha$. In the prediction period, some of the observers from the set of $\alpha$s may be missing (they did not observe). Similarly, there might be observations provided by other observer, for which the coefficients from the previous evaluation period do not exist (she/he did not observe during the evaluation period). Such an observer is not considered in forming the composite within the given prediction window.} we obtain a prediction based on this observer, $g^\beta_0$ and $f^\beta_0$:
\begin{equation}
    g^\beta_0(D)=k^\beta_g g^\beta(D)\quad{\rm and}\quad f^\beta_0(D)=k^\beta_f f^\beta(D).
\end{equation}
The composite prediction $g_0(D)$ and $f_0(D)$ is computed as a weighted average over observers $\beta$, where the weights correspond to the number of observations $M^\beta$ that served to determine the conversion coefficients in the evaluation period:
\begin{equation}
    g_0(D)=\sum\limits_\beta w^\beta g^\beta_0(D)\quad{\rm and}\quad f_0(D)=\sum\limits_\beta w^\beta f^\beta_0(D),
\end{equation}
with
\begin{equation}
    w^\beta=\frac{M^\beta}{\sum_\beta M^\beta}. \label{eq:weights}
\end{equation}
As a final step before storing the composite prediction to target we round-off $g_0$ and $f_0$ to integers and compute $R_0(D)=10g_0(D)+f_0(D)$. The rounded predicted composites for a set of days $D$ within the window of length $\Delta W$ are added to the target series \emph{after} the evaluation window of length $W$ in the direction of time arrow. The days from the prediction window, when no observations are available, remain unfilled. 

Next we slide the evaluation window by $\Delta W$ forward in time and repeat the process. This adds a new segment having a length of $\Delta W$ to the end of the target series. We slide again by $\Delta W$ until we reach the end of the series of available observations. We refer to the progress following the arrow of time as the ``half-iteration''. 

To complete the ``iteration'' we now switch the progress against the arrow of time and start from the end of the series. All the values of the target series are dropped except for the values located in the window of $W$ days at the very end of the series\footnote{Note that this target series cropping also drops all the values from the initial series.}. We then repeat the procedure described above against the arrow of time by subsequent sliding by $\Delta W$ until we reach the beginning of the series. At each step a new segment of the target series is built \emph{in front} (in the direction of the time arrow) of the target series. That is, a scheme displayed in Fig.~\ref{fig:method} is mirrored in the horizontal direction. The progress is stopped when the beginning of the available observations is reached. The progression against the arrow of time makes the target series denser because it allows to fill also days, where the common observations were not available in the first half-iteration. 

If necessary, the process may go iteratively forth and back again. In that case, the process starts from the $W$-days long window of the target series from the previous iteration, all the other values of the target series beyond this starting window are dropped. The target series is then built again by segments from the start without a direct connection to the previous realisation (except for the starting $W$-days long window).

\section{Testing the methodology}

Our proposed methodology contains several non-trivial steps that need be justified by testing. 

\begin{figure}
    \centering
    \includegraphics[width=\textwidth]{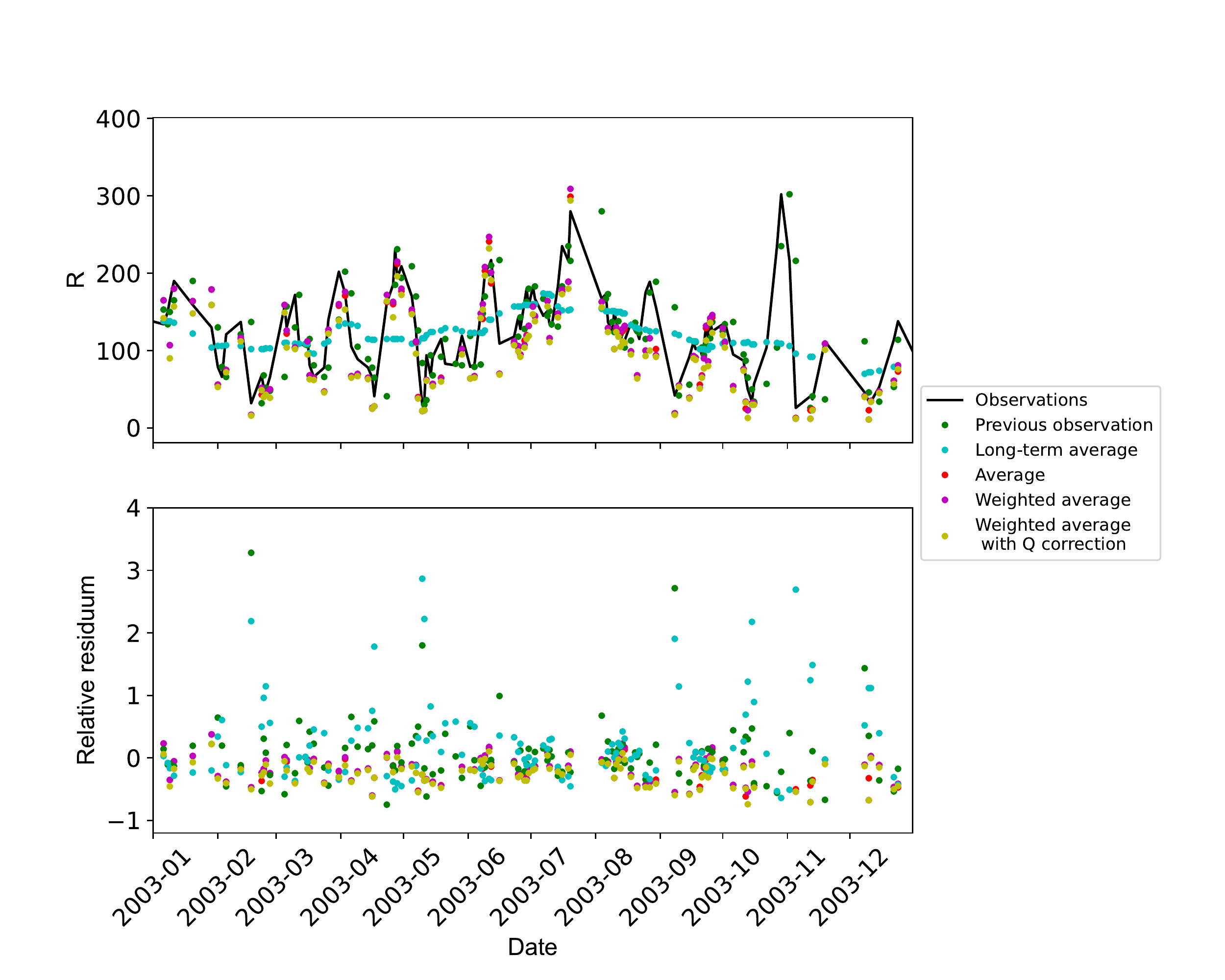}
    \caption{In the upper panel one can see the observations of the tested observer (F.~Zloch) as compared with their predictions by using various tested models. In the bottom panel, the corresponding relative residua are shown. Only one year 2003 from the validation period spanning over 14 years is displayed.}
    \label{fig:residua}
\end{figure}

\subsection{Validation of the principle}

In the prediction step, the choice of the weighted average with the reduction of observations to optimal observing conditions is not obvious. The choice is based on testing of various predictive models. 

The tests were performed separately again for 8 observers with the largest number of observations in the data we have at disposal. The testing followed the principles usual in machine-learning methods. For each observer considered (termed the \emph{tested} observer henceforth), we split his/her observations in two equally populated halves in time. The first half mimicked the evaluation period of the algorithm described in the previous section, whereas the second half mimicked the prediction period and served as validation of the predictive model. 

Following the reduced version of the methodology, we computed personal conversion coefficients of other observers with respect to the tested observer in the first half of his/her recordings. Then we used a set of predictive models to predict what should the tested observer record in the second half. The existence of the real tested observer's observations during the second half allowed us to compare the predictions with observations and to determine the success rates of various predictive model. 

In the evaluation, we considered a following set of predictive models: 
\begin{itemize}
    \item[\emph{Previous observation}] This is a simple model which does not involve other observers, however, it is used in testing of machine-learning methods. In the validation period, we considered the previous available observation of the tested observer as the prediction. 
    \item[\emph{Long-term average}] Again, this is a simplified model that does not involve other observers, but it again is used in testing the machine-learning methods. The prediction is based on the average of observations of the tested observer over previous 30 days. 
    \item[\emph{Average}] For the given day, the observations of other available observers are scaled by the appropriate coefficients $k^\alpha_g$ and $k^\alpha_f$ and a arithmetic mean over all of these observers is considered. The personal conversion coefficients were computed by using raw non-reduced observations, i.e. with $c_g=0$ and $c_f=0$ in (\ref{eq:gred}) and (\ref{eq:fred}).
    \item[\emph{Weighted average}] This predictive model is similar to previous, only weighting given by (\ref{eq:weights}) is applied. The logic behind weighting says that when more pairs during the evaluation period are available, the derived coefficients are expected to be more robust. 
    \item[\emph{Weighted average with Q correction}] This predictive model is again an advance of the previous. In the evaluation period both the observations of the tested observer and the other observers are reduced to observing conditions with $Q=5$ by using (\ref{eq:gred}) and (\ref{eq:fred}).
\end{itemize}

\begin{table}[]
    \centering
    \begin{tabular}{r|cccccc}
\toprule
model & $RB(f)$ & $RMSE(f)$ & $RB(g)$ & $RMSE(g)$ & $RB(R)$ & $RMSE(R)$ \\
\midrule
Previous observation & $0.087$ & $0.262$ & $0.321$ & $3.354$ & $0.095$ & $0.334$ \\
Long-term average & $0.109$ & $0.263$ & $0.582$ & $4.896$ & $0.118$ & $0.383$ \\
Average & $-0.098$ & $0.062$ & $-0.125$ & $0.197$ & $-0.114$ & $0.073$ \\ 
Weighted average & $-0.102$ & $0.061$ & $-0.111$ & $0.206$ & $-0.110$ & $0.075$\\
Weighted average, & $-0.136$ & $0.068$ & $-0.206$ & $0.200$ & $-0.170$ & $0.082$ \\
Q correction & \\
\bottomrule
\toprule
model & $min_{\delta f}$ & $max_{\delta f}$ & $min_{\delta g}$ & $max_{\delta g}$ & $min_{\delta R}$ & $max_{\delta R}$ \\
\midrule
Previous observation & $-0.80$ & $4.00$ & $-0.923$ & $32.000$ & $-0.818$ & $7.308$ \\
Long-term average & $-0.75$ & $4.00$ & $-0.947$ & $41.000$ & $-0.974$ & $6.454$ \\
Average & $-0.75$ & $1.00$ & $-0.947$ & $3.000$ & $-0.814$ & $1.290$ \\ 
Weighted average & $-0.75$ & $1.00$ & $-0.950$ & $3.000$ & $-0.814$ & $1.290$\\
Weighted average, & $-0.75$ & $1.00$ & $-0.950$ & $3.000$ & $-0.814$ & $1.182$ \\
Q correction & \\
\bottomrule

    \end{tabular}
    \caption{Example statistical evaluation of various considered predictive models, here summarised for F.~Zloch over the validation period of 14 years. }
    \label{tab:table_models}
\end{table}

\noindent To assess the success rate of various models we employ the visual comparison (such as that displayed in Fig.~\ref{fig:residua} for observer František Zloch) and following four statistical quantities. For each predictive model we evaluated the relative mean bias $RB$
\begin{equation}
    RB(t) = \frac{1}{N} \sum\limits_i^N (t_{i,\rm m} - t_{i,\rm o})/t_{i,\rm o}, \label{eq:RE}
\end{equation}
where $t$ is the tested value $t=\{g,f,R\}$, $N$ is the number of validated pairs, $t_{i, \rm m}$ is the model prediction, $t_{i, \rm o}$ is the observation. Similarly we defined the relative mean squared error
\begin{equation}
    RMSE(t) = \frac{1}{N} \sum\limits_i^N \left[ (t_{i,\rm m} - t_{i,\rm o})/t_{i,\rm o} \right]^2. \label{eq:RMSE}
\end{equation}
As auxiliary criteria we also used maxima and minima of the relative residua,
\begin{equation}
    min_{\delta t} = \min\limits_{i \in 1..N} \left[ (t_{i,\rm m} - t_{i,\rm o})/t_{i,\rm o} \right] \quad {\rm and} \label{eq:min_res}
\end{equation}
\begin{equation}
    max_{\delta t} = \max\limits_{i \in 1..N} \left[ (t_{i,\rm m} - t_{i,\rm o})/t_{i,\rm o} \right]. \label{eq:max_res}
\end{equation}

The quantities corresponding to the example in Fig.~\ref{fig:residua} are given in Table~\ref{tab:table_models}. Situation for the other 7 observers is very similar, hence the conclusions drawn from the example of František Zloch are general. 

We see that the model of previous observation and long-term average have the largest residuals and from the tested predictive models perform the worst. This could be expected for the long-term average model, because when using a 30-day average it cannot reproduce the day-to-day changes, however it is somewhat surprising that it does not perform significantly worse than the previous observation model. The typical time scale for significant changes on the surface of the Sun is a few days. On the other hand, it cannot be guaranteed that the previous observation was always in the previous day, which would assure a gradual change. We find both these models unsuitable. 

The other models on the other hand seem to perform comparably. The only statistical quantity where the model \emph{weighted average with Q correction} performs slightly better is the maximum of the relative residuum. It would seem that inclusion of the reduction to the fixed observing conditions helps to suppress the outliers.

Following these tests we gave the preference of the latest predictive model over the others. Lastly, we would like to point out that in these tests we did not allow for changes of the values of personal conversion coefficients in time. They do evolve, however, and the neglection of this effect might increase the level of deviations reproduced in the assessed statistical quantities. 

\subsection{Long-term trends in observers' coefficients}

Previous sections formed the principles of our methodology that allowed to finalise the computer code \citep{THECODE}. It is written in {\sc Python} when using {\sc Numpy} \citep{harris2020array} and {\sc Pandas} \citep{mckinney2010data} packages. Especially the later one allowed us to write a flexible and universal code. What is missing is to determine the optimal lengths of evaluation window $W$, prediction window $\Delta W$ and number of iterations. 

To set the reasonable value of $W$ we need to look at the evolution of the conversion coefficients. We computed the personal conversion coefficients of each observer in our dataset with respect to the reference series, which was constructed as a simple mean of the available observations and then averaged over 28~days (one solar rotation). The coefficients were computed in a sliding window having $W=1000$~days, which is the shortest reasonable given the scarcity of our dataset. Shorter windows may be possible, however our dataset is not dense enough so that in shorter windows the coefficients are not representative as only a small number of pairs entering (\ref{eq:k_W}) is available. 

\begin{figure}
    \centering
    \includegraphics[width=\textwidth]{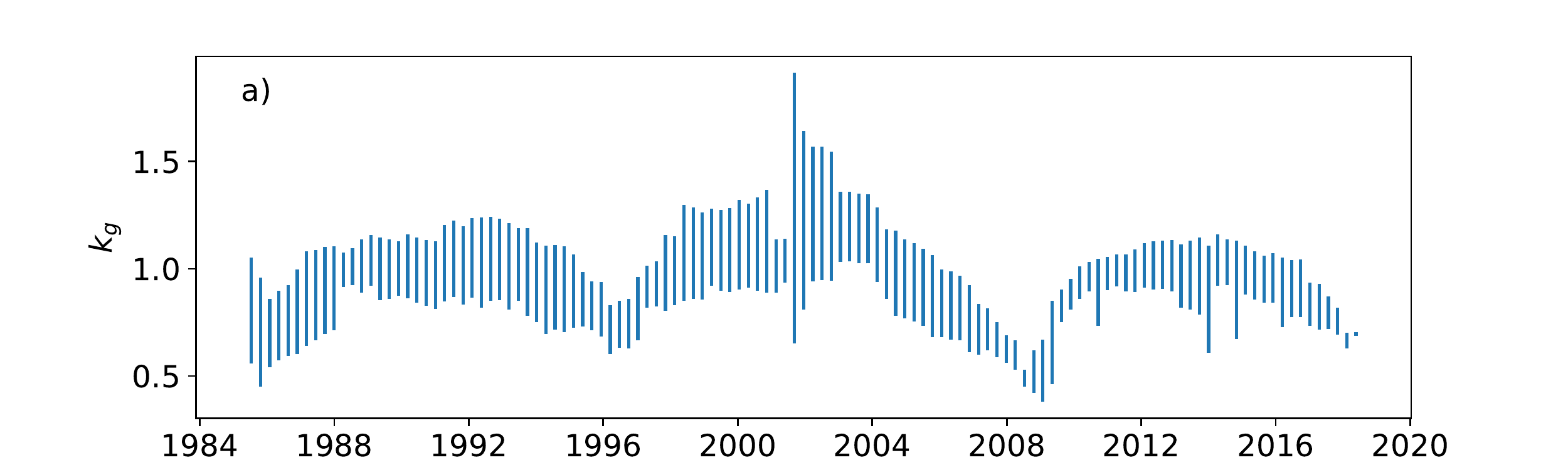}
    \includegraphics[width=\textwidth]{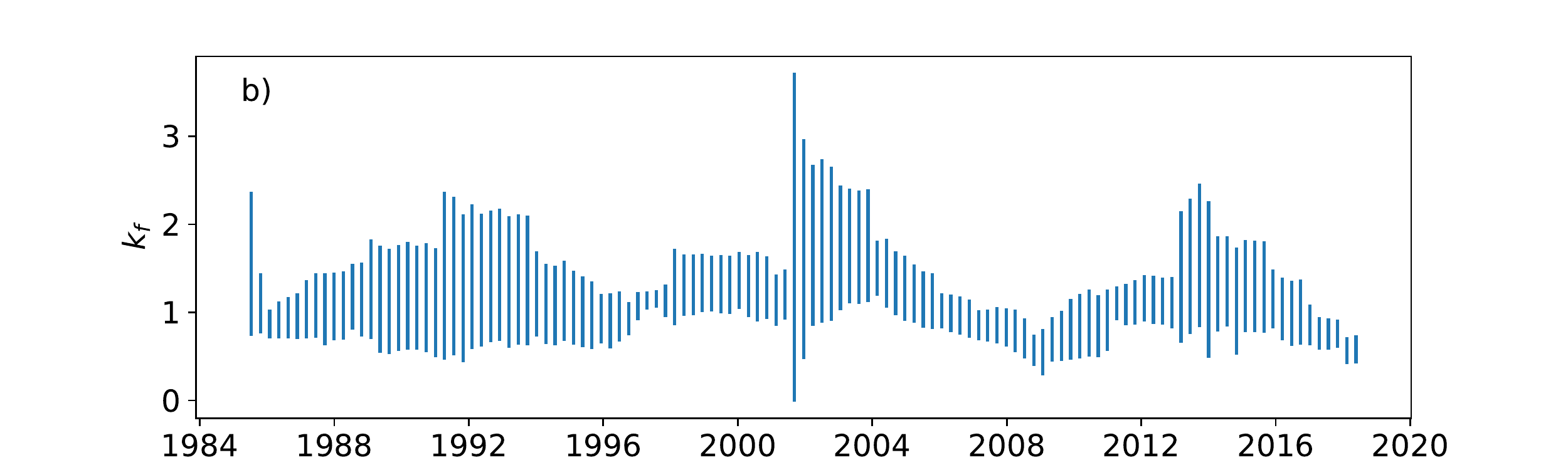}
    \caption{1-$\sigma$ ranges of the conversion coefficients of the observers with respect to the simple mean over the studied period. Both group number ($g$, panel a) and sunspot count ($f$, panel b) are plotted. Obtained with $W=1000$~days. }
    \label{fig:mean_ks}
\end{figure}

Then we plotted the statistics of the conversion coefficients of all available observers in Fig.~\ref{fig:mean_ks}. The bars are centered on the mean value of the conversion coefficients of all observers within the considered window and extend 1$\sigma$ to either side. Interestingly enough, the coefficients as compared to the mean observations seem to be correlated with the solar cycle, that is they \emph{on average} seem to increase during the maxima and decrease during the minima. It is especially prominent in the case of $k_g$. 

We may only speculate what is the cause of this trend. During the maxima, the observers are overwhelmed by an overall increased activity, so that some of them easily miss for instance the Axx-type groups (in \citet{1990SoPh..125..251M} classification it represents an isolated pore). Also, possibly, the splitting of the complicated sunspots nests into an appropriate number of magnetically coupled groups may be difficult for visual observers. On the other hand, during the minima the observers pay a special attention when scanning the solar disc and hence find even smallest Axx-type groups. 

The long-term evolution of the personal conversion coefficients sets limits on an appropriate selection of length of $W$, which should be longer than a solar cycle. The values of $W$ shorter than a cycle may introduce rising/declining trends in the conversion coefficients that affect the overall determination of $R_0$. Too short $W$s thus lead to an inappropriate evaluation of the coefficients, where during the consecutive windows shifted by $\Delta W$ one can see a positive feedback leading to the incorrect solution. 

The above described hypothesis based on the rules-of-thumb may be tested by exploring a large $(W,\Delta W)$ space. We performed the execution of our code for a set of $W$s and $\Delta W$s and studied the effects of the choice on the two important quantities.

\begin{figure}
    \centering
    \includegraphics[width=0.49\textwidth]{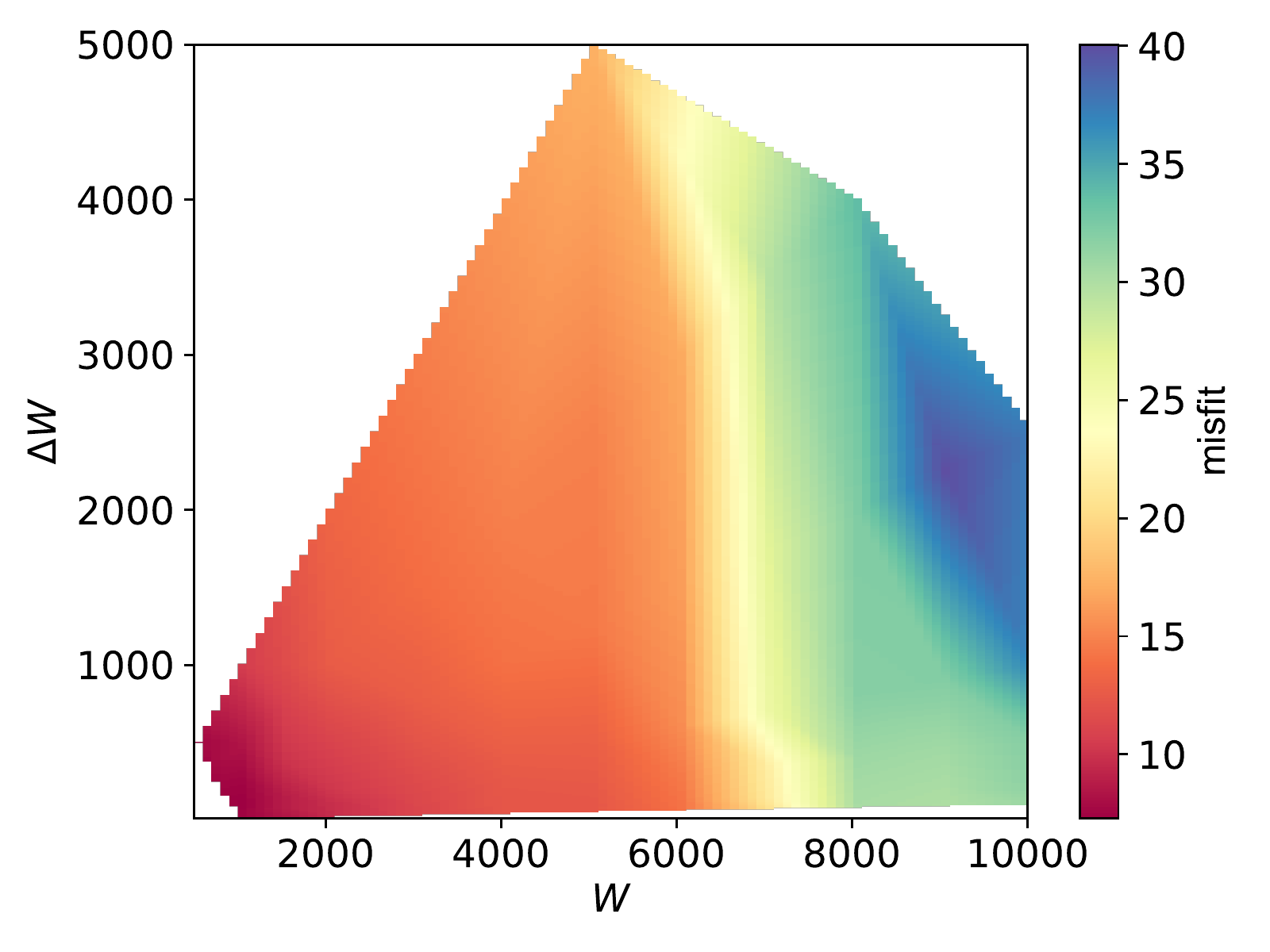}
    \includegraphics[width=0.49\textwidth]{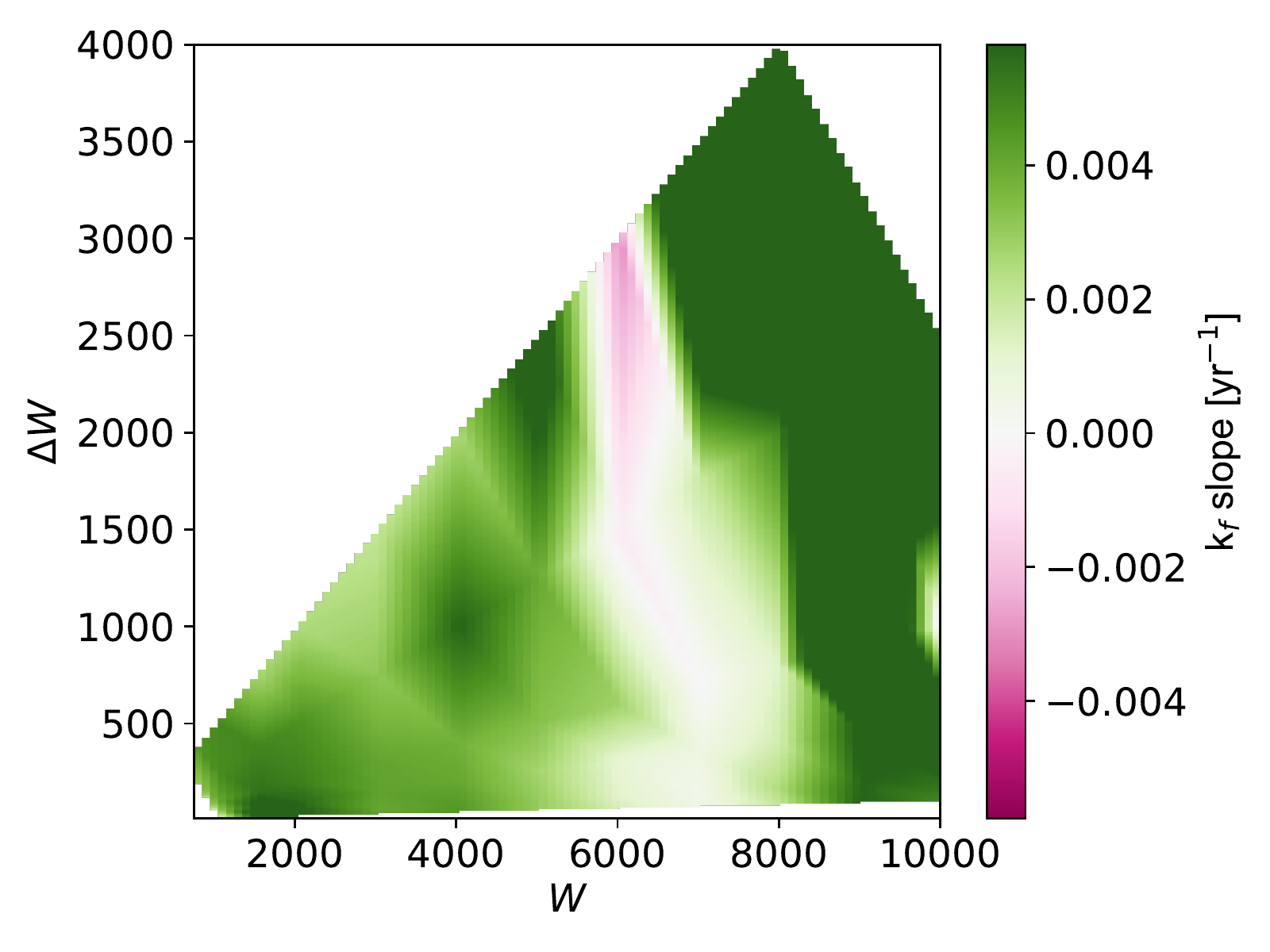}
    \caption{Maps of the auxiliary criteria helping to determine optimal values of $W$ and $\Delta W$. On the left a map of misfits (\ref{eq:misfit}) are plotted, on the right we plot the slope of the linear trend of the average personal scaling coefficient $k_f$. }
    \label{fig:WdW_space}
\end{figure}

First we introduced a \emph{misfit}. The misfit described how well does the target series represent the actual records of the individual observers. The misfit is then defined as
\begin{equation}
    {\rm misfit} = \left[ \frac{1}{N_D} \sum\limits_D \frac{\sum\limits_\beta \left( R_0(D)- [10k_g^\beta(D)g^\beta(D)+k_f^\beta(D)f^\beta(D)] \right)^2}{\sum_\beta M_\beta} \right]^{1/2}.
    \label{eq:misfit}
\end{equation}
The misfit in fact evaluates the mean squared deviation of the target series and $k$-coefficient transformed values of individual observers, evaluated for all considered observers $\beta$ and in all days $D$. The values of the misfit are displayed in Fig.~\ref{fig:WdW_space} in the left panel. The misfit is very small for small $W$s, then there is a plateau between say $W=3000$~days and $W=6000$~days and for larger values of $W$ the misfit gradually increases. We interpret the figure in such a way that for $W<3000$~days the methodology fits the observations very well, because it naturally allows for fast changes of the personal scaling coefficients. For $W>6000$~days the misfit increases because the personal coefficients vary on shorter time scales. In other words, for $W<3000$~days the method overfits the data, whereas for $W>6000$~days it underfits the data. The reasonable values of $W$ then naturally lie within the plateau between 3000 and 6000~days. The dependence on $\Delta W$ is much weaker, it would seem that within the $W$-plateau $\Delta W<3000$~days does not affect the misfit significantly. 

Another quantity we consider is the overall trend in the personal conversion coefficients. For each day we compute the mean personal conversion coefficients for observers who provided their records on that day and its standard deviation. Then we use a linear fit through the mean personal coefficients with standard deviations as measurement errors in time. This allows us to obtain a slope of a general linear trend in the personal scaling coefficients. The fit is performed independently for $k_g$ and $k_f$, however, $k_f$ shows a stronger dependence on $W$ and $\Delta W$ and hence it is considered further. In general, one expects the slope of the overall trend in the scaling coefficients to be around zero. Non-zero trends indicate a systematic problem either with the target series (the solution) or with a large portion of the observations. Naturally, non-linear, e.g. oscillating, trends cannot be excluded and such trends in principle could escape detection by the linear fitting. Due to the two-step nature of our algorithm the systematic deviations of the personal scaling coefficients multiply during the construction of the target series and usually end up with the long-term systematic trend that would be revealed by the linear fit. 

The map of the $k_f$ trend is plotted in Fig.~\ref{fig:WdW_space} in the right panel. According to this the optimal values of $W$ lie around 6000~days, where it is better to use a rather small value of $\Delta W$. 

\begin{figure}
    \centering
    \includegraphics[width=\textwidth]{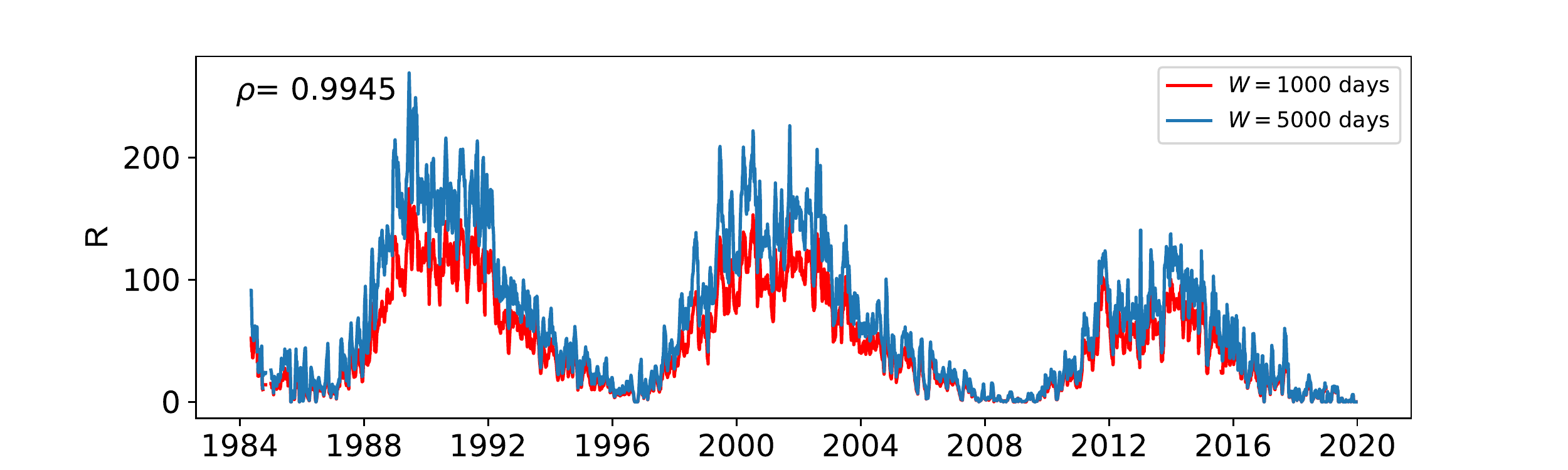}\\
    \includegraphics[width=\textwidth]{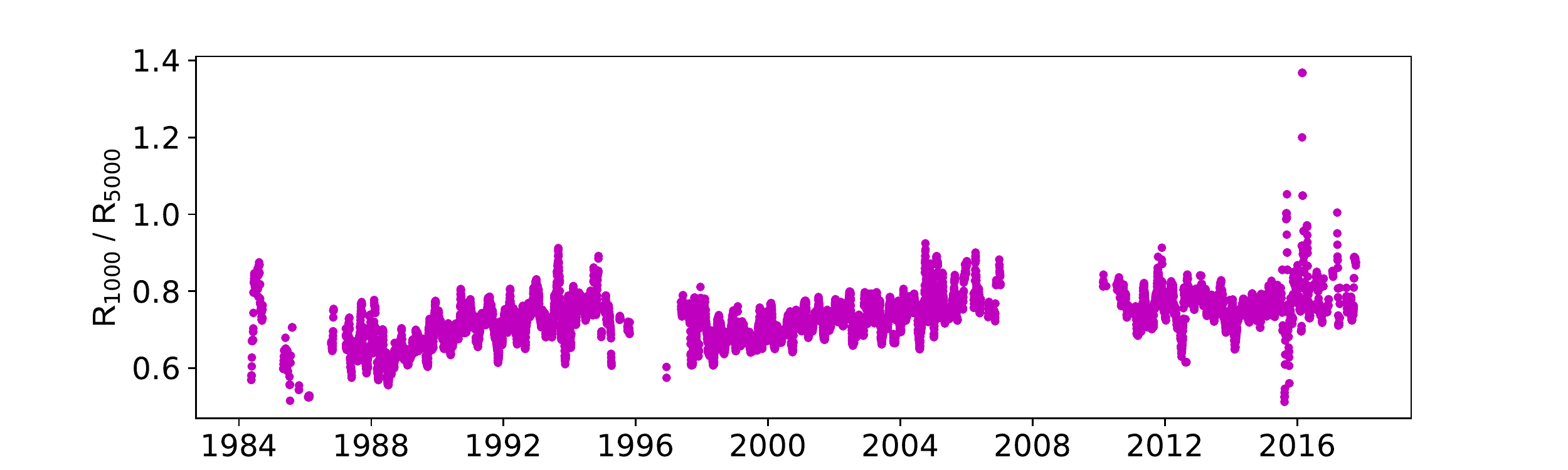}
    \caption{Two different solutions of $R$ by our methodology obtained with $W=1000$~days and $W=5000$~days (the fiducial solution), both with $\Delta W=100$~days. Obviously, both solutions are highly correlated, however the solution with a smaller $W$ shows a clear envelope trend (the overall amplitude of $R$ is considerably smaller during cycle 23 and 24. This envelope trend is seen in the lower panel, where the ratio of two solutions from the upper panel is plotted. This behaviour is due to the trend in the conversion coefficients that were evaluated on the scales shorter than the cycle-long oscillation.}
    \label{fig:W_effect}
\end{figure}

This may easily be demonstrated by comparing the results of runs differing only by choice of $W$, as plotted in Fig.~\ref{fig:W_effect}. There a solution with a short $W=1000$~days differs significantly in the overall amplitude from the solution with $W=5000$~days, whereas the overall structure of the resulting target series are very close (with a Pearson's correlation coefficient close to unity). We tested that the overall increasing or decreasing trend in the amplitude of the resulting $R_0$ is indeed due to the trends in the personal conversion coefficients, as for such a solution plots similar to that presented in Fig.~\ref{fig:mean_ks} clearly show an almost steady decrease or increase. 

\begin{figure}
    \centering
    \includegraphics[width=\textwidth]{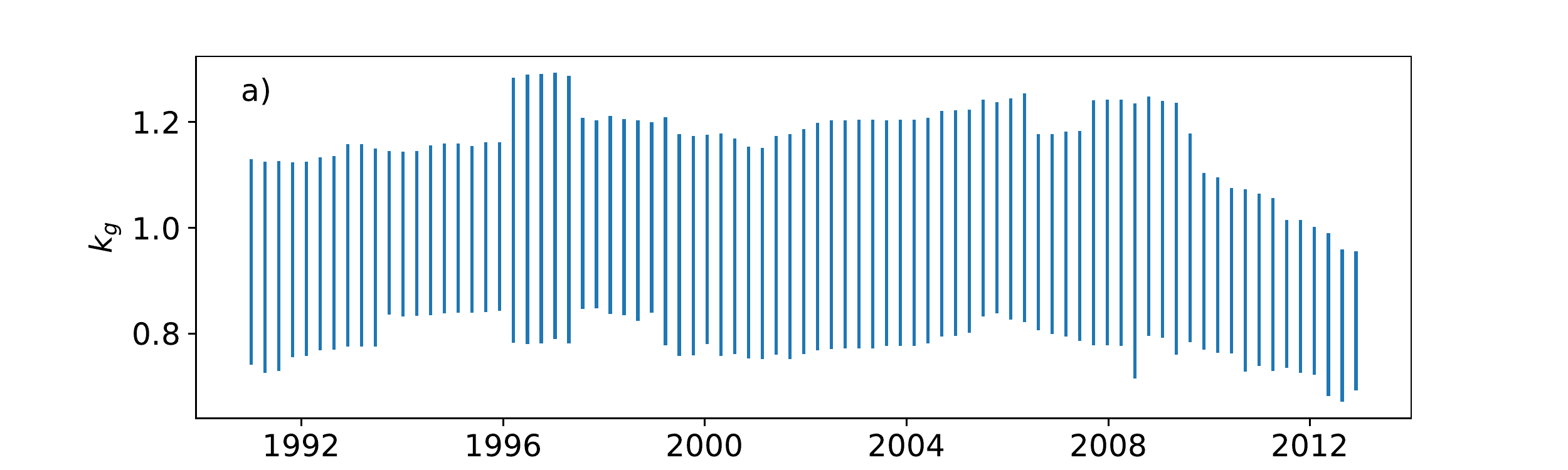}
    \includegraphics[width=\textwidth]{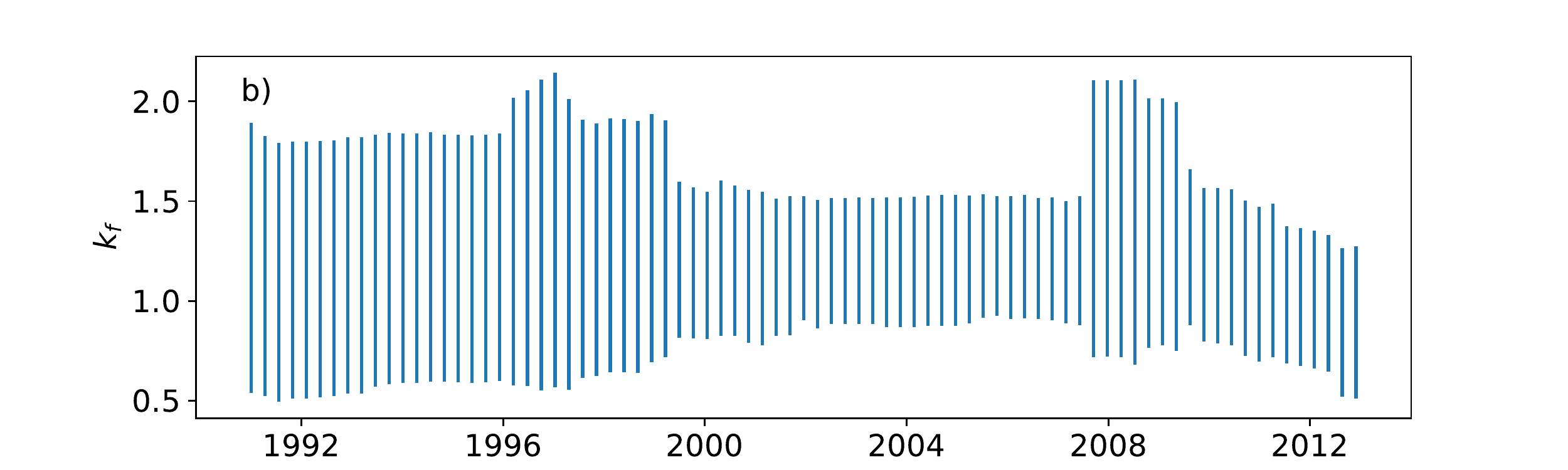}
    \caption{Same as Fig.~\ref{fig:mean_ks}, only these plots were obtained with $W=5000$~days. }
    \label{fig:mean_ks_5000}
\end{figure}

A choice of an appropriate $W$ and $\Delta W$ is thus not straightforward. One should pay attention to appropriately assess the properties of the resulting series and evaluate critically the auxiliary criteria. Such a criterion is again a plot like that in Fig.~\ref{fig:mean_ks}, where a gradually increasing or decreasing trend usually indicates an inappropriate (too short) choice of $W$. Also coefficients of individual observers are to be studied. For instance, when a personal coefficient of an observer changes by more than 50\% over few years, this indicates a serious issue with the observations of this observer. Should this kind of rapid change be seen at several observers, that on the other hand is an indication of an inappropriate selection of $W$. In our case, we found a choice of $W=5000$~days a good trade-off between the suppression of the cycle-related changes and time resolution (compare Figs.~\ref{fig:mean_ks} and \ref{fig:mean_ks_5000}, where cycle-related changes are suppressed). 

Fig.~\ref{fig:WdW_space} forces us to use a rather small value of $\Delta W$ for the selected $W=5000$~days. There are, however, again other considerations. With a larger $\Delta W$ a temporal resolution in determination of the coefficients is lost and one may then miss e.g. an unexpected trend is someone's coefficients. On the other hand, the choice of $\Delta W$ too short pose a risk of not finding enough suitable observers in the prediction period, which pose a risk of interruption in building the target series by adding an empty segment of the length of $\Delta W$ to the end. This usually occurs for short $\Delta W$ during the deep and long minimum, because only observations where $g>0$ and $f>5$ are considered. Similar situation occurs when there is a long gap in observations. In the case when these empty segments accumulate, the building does not recover after passing the minimum or the gap and the code run leads to a valid but shortened target series. In our case, we found an optimal choice $\Delta W=100$~days. 

\subsection{Number of iterations}
As we indicated in the description of the calculation algorithm, in principle, the process is iterative. It is not clear from the beginning as to how many iterations are necessary in order to obtain a stable solution or even if the iterative procedure converges. With the code, it is easy to test it, however, many iterations may be lengthy. Using our set of about 16\,000 individual observations one full iteration takes a few minutes running single-thread on M1-based Macbook Air. 

\begin{figure}
    \centering
    \includegraphics[width=0.49\textwidth]{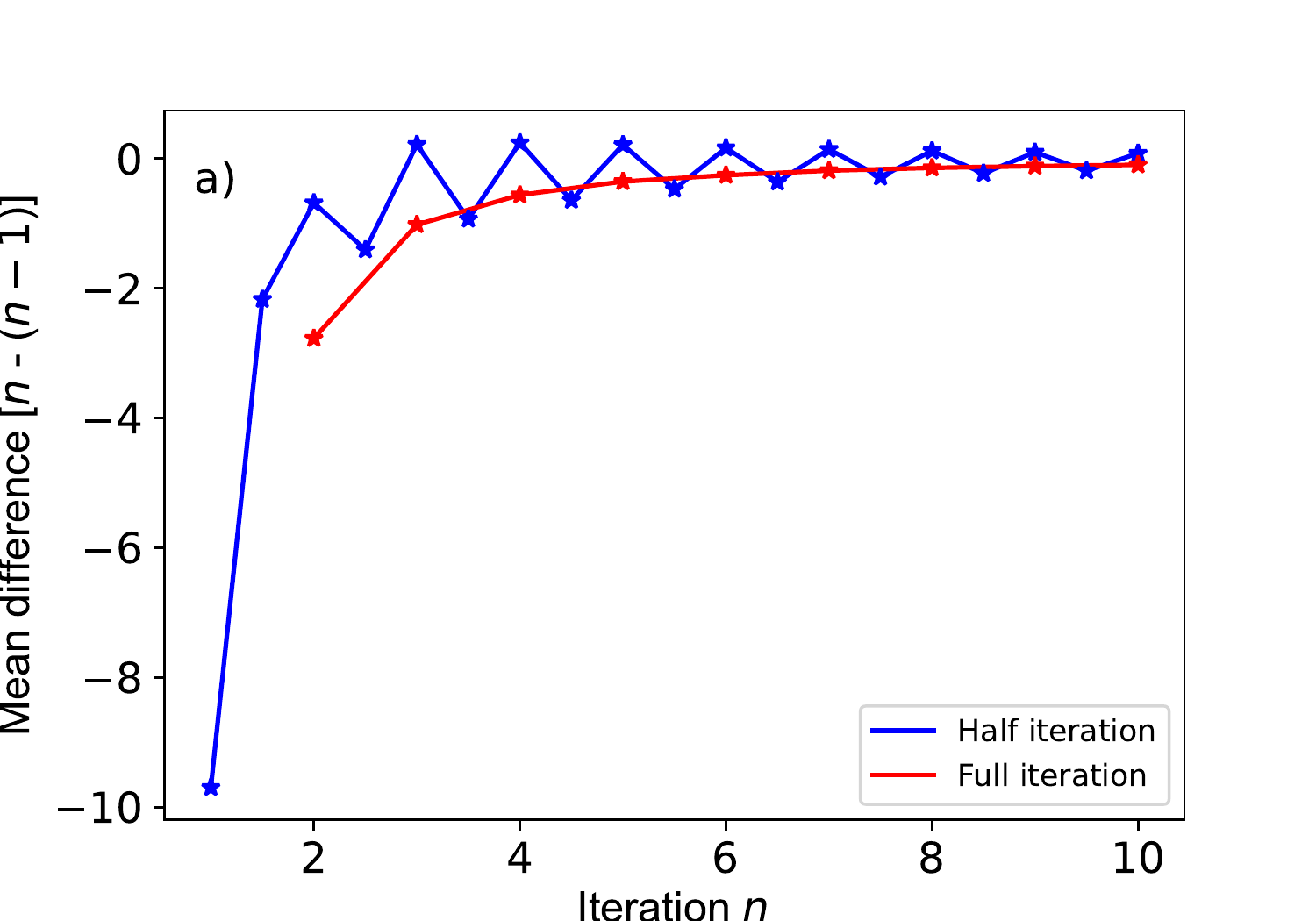}
    \includegraphics[width=0.49\textwidth]{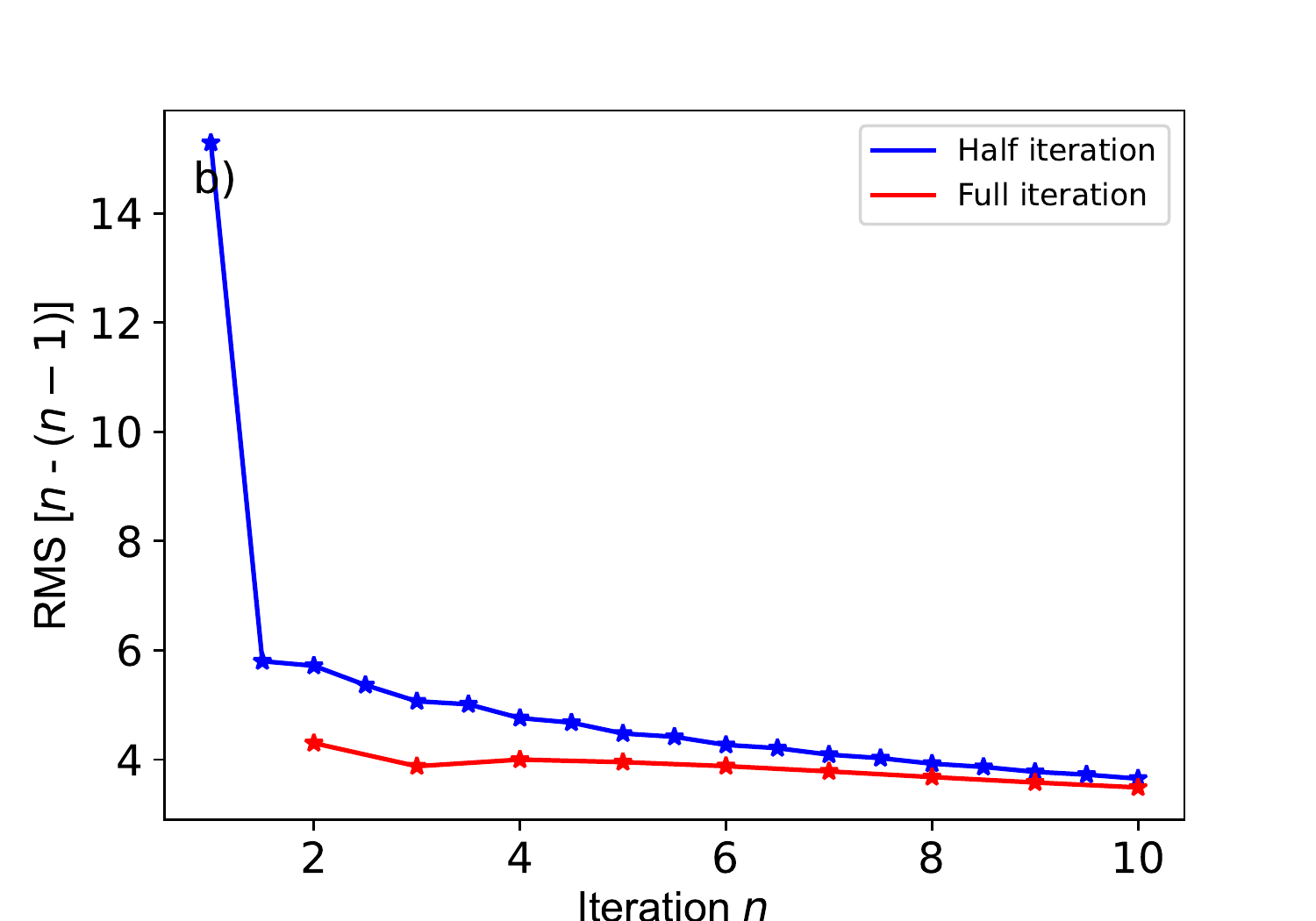}
    \caption{Convergence of the data series after $n$ iterations. a -- The mean difference of the resulting $R$ obtained for the $n$-th and $(n-1)$-th iteration. b -- RMS of the differences. }
    \label{fig:data_iteration_convergence}
\end{figure}

Statistical quantities defined by equations (\ref{eq:RE}) and (\ref{eq:RMSE}) provide us with suitable quantities that could be used to study the convergence of the iterations. Therefore, using analogous formulae, we evaluate the mean difference and the mean squared difference of the computed $R_0$ between consecutive iterations. Both quantities are plotted in Fig.~\ref{fig:data_iteration_convergence} for the solution with $W=5000$~days and $\Delta W=100$~days. 

To properly interpret Fig.~\ref{fig:data_iteration_convergence} we remind that the term ``half-iteration'' indicates the progression through the data series respecting the time arrow, ``full iteration'' includes also the backward propagation against the arrow of time. Each point in the plot then gives the overall $RB$ comparing the results from the previous half/full iteration. By looking at panel a), one would perhaps naively choose e.g. the point at 3 iterations, where the $RB$ seems considerably smaller than for 2 iterations. However, this means that the 3rd iteration is very similar to the 2nd iteration and hence the 3rd iteration does not bring forward any significant improvement. In the panel b) it is even more evident that 2 iterations are sufficient. 

Putting all the above given arguments together, our \emph{fiducial solution} respecting optimally the properties of the archive of observations we are processing has thus the following choice of free parameters: The width of the evaluation window $W=5000$~days, the width of the prediction window $\Delta W=100$~days, 2~full iterations. 

\subsection{Sensitivity to the choice of the initial series}
\label{sect:initial}
Our methodology is iterative, therefore there must exist an initial guess of the reference series, which is used for the evaluation of the weights in the first evaluation windows. A successful methodology should be rather robust with respect to the utilisation of the initial series. 

In the paragraphs above, we pointed out that a reasonable choice of the initial series is anything ``solar-like'', that is, it should somewhat resemble the expected solution in term of the general trends. One obvious choice is to use the observations of one respected observer, experienced enough and with a large set of observations available. From the methodology it follows that this observer should provide us with observations that span over a period at least $W$ long, otherwise it will not be possible to compute the personal conversion coefficients of the other observers. Professional observers or devoted amateurs usually fulfill this condition. In our case we chose the observers working for the Solar patrol at Ond\v{r}ejov observatory, namely Franti\v{s}ek Zloch. From the Czech amateurs we certainly must mention Ladislav Schmied, who in his private observatory drew the Sun for 66 years having altogether more then 12\,500 drawings. Unfortunately, not all drawings by Ladislav Schmied were digitised yet. 

Another choice would be some sort of the aggregated series. A choice which offers itself automatically is a simple arithmetic average of all observations in the series. Such a choice ensures that the general solar trends (such as the 11-year cycle) will be respected and also it ensures that for each considered day there will be at least one observer for which one may determine the conversion coefficients. 

\begin{table}[]
    \centering
    \emph{Correlation coefficients}

\begin{tabular}{lrrrr}
\toprule
{} &  Zloch &  Schmied &   Ivan &  average \\
\midrule
Zloch   & 1.0000 &   0.9987 & 0.9998 &   0.9998 \\
Schmied & 0.9987 &   1.0000 & 0.9983 &   0.9987 \\
Ivan    & 0.9998 &   0.9983 & 1.0000 &   0.9999 \\
average & 0.9998 &   0.9987 & 0.9999 &   1.0000 \\
\bottomrule
\end{tabular}
\vskip 6pt 
\emph{Slopes}

\begin{tabular}{lrrrr}
\toprule
{} &  Zloch &  Schmied &   Ivan &  average \\
\midrule
Zloch   & 1.0000 &   0.7637 & 1.0390 &   0.9726 \\
Schmied & 1.3078 &   1.0000 & 1.3586 &   1.2721 \\
Ivan    & 0.9623 &   0.7349 & 1.0000 &   0.9361 \\
average & 1.0280 &   0.7851 & 1.0682 &   1.0000 \\
\bottomrule
\end{tabular}
    \caption{Mutual correlation coefficients and multiplicative coefficients of final sunspot number when using different initial observations. }
    \label{tab:different_initial}
\end{table} 

Our tests show that the results only weakly depend on the choice of the initial series in terms of the evolution in time. This is demonstrated in Table~\ref{tab:different_initial}. Here we evaluated two statistical quantities comparing the results of the fiducial runs with different initial series. First we compute the Pearson's correlation coefficient, second we computed the quantity we refer to as \emph{slope}. The slope corresponds to the optimal multiplicative factor between two series in the least-squares sense. It basically gives the slope of the linear fit without a constant term. Table~\ref{tab:different_initial} clearly shows that the resulting series does not depend on the initial series in terms of the evolution in time, that is, the correlation coefficients are close to unity. However, due to the different conversion coefficient of the initial series the overall amplitude may differ. That is, the resulting $R_0$ series computed from different initial series are identical except for the multiplicative scaling. 

\begin{figure}
    \centering
    \includegraphics[width=\textwidth]{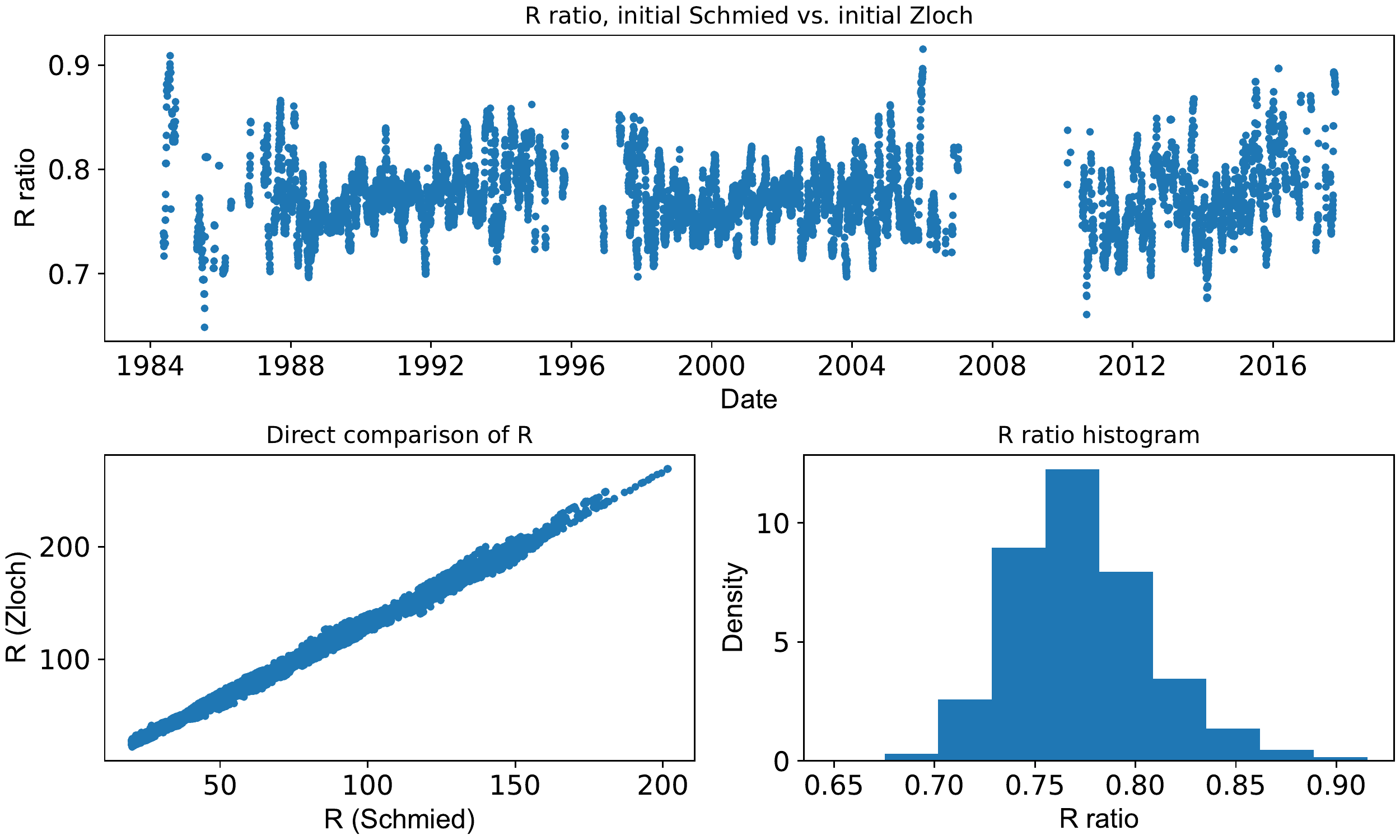}
    \caption{Direct comparison of the sunspot number resulting from our code with the initial series coming from F.~Zloch and L.~Schmied. In the upper panel, the ratio of the solutions is plotted with respect of time. In the bottom-left panel the corresponding relative numbers are plotted against each other, in the bottom-right panel one can see the histogram of the ratios plotted in the upper panel.}
    \label{fig:initials_Zloch_Schmied}
\end{figure}

The scaling is indeed multiplicative only, as shown e.g. in Fig.~\ref{fig:initials_Zloch_Schmied}, where the solutions starting from observer F.~Zloch is directly compared with the solution starting from L.~Schmied (note that this pair of observers has a large scaling factor -- cf. Table~\ref{tab:different_initial}). The ratios of both solutions vary around the slope value (which is $0.7637$ in this case), the histogram indicates the relative residua being single-modal and its shape is what one expects from the multiplicative model. The mutual scatter plot also does not indicate any pathology. 

The multiplicative scaling found for various initial series is a natural property of our methodology. As we pointed out, the methodology aims to construct a target series of observations by a hypothetical stable observer who combined \emph{the best of all}. The choice of the initial series thus influences the overall magnitude of the target series, it influences the ``personal coefficients'' of the hypothetical observer. Real observers are known to have different scaling coefficients and our is not an exception. Besides, various definitions of relative sunspot numbers used in the world have or had overall different magnitudes, as already discussed in the Introduction. 

\begin{figure}
    \centering
    \includegraphics[width=\textwidth]{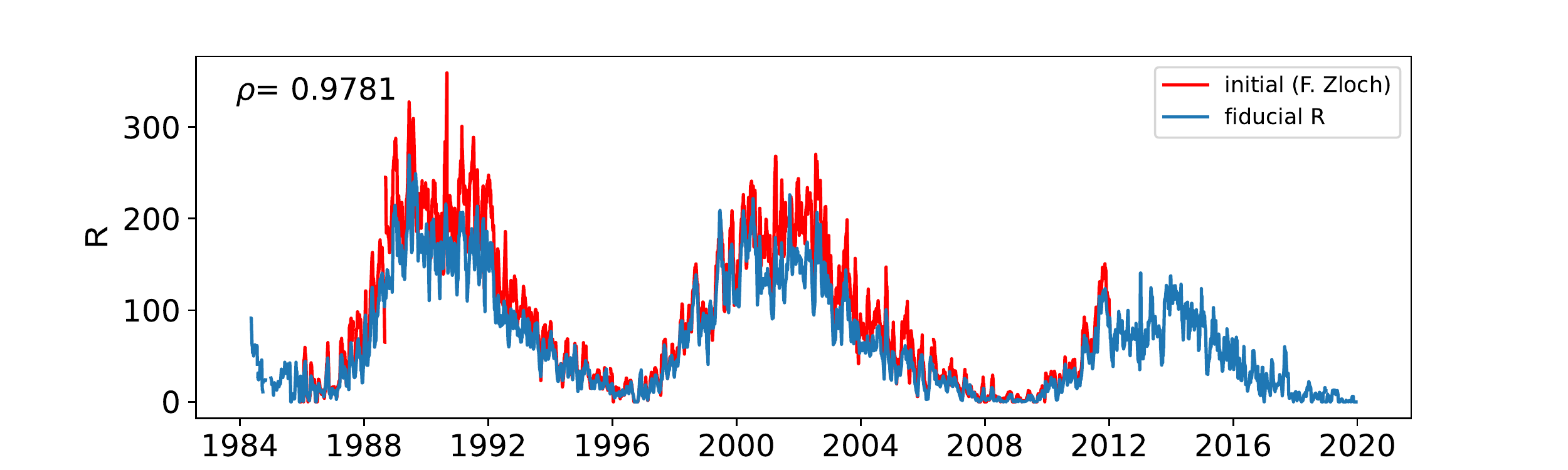}
    \includegraphics[width=\textwidth]{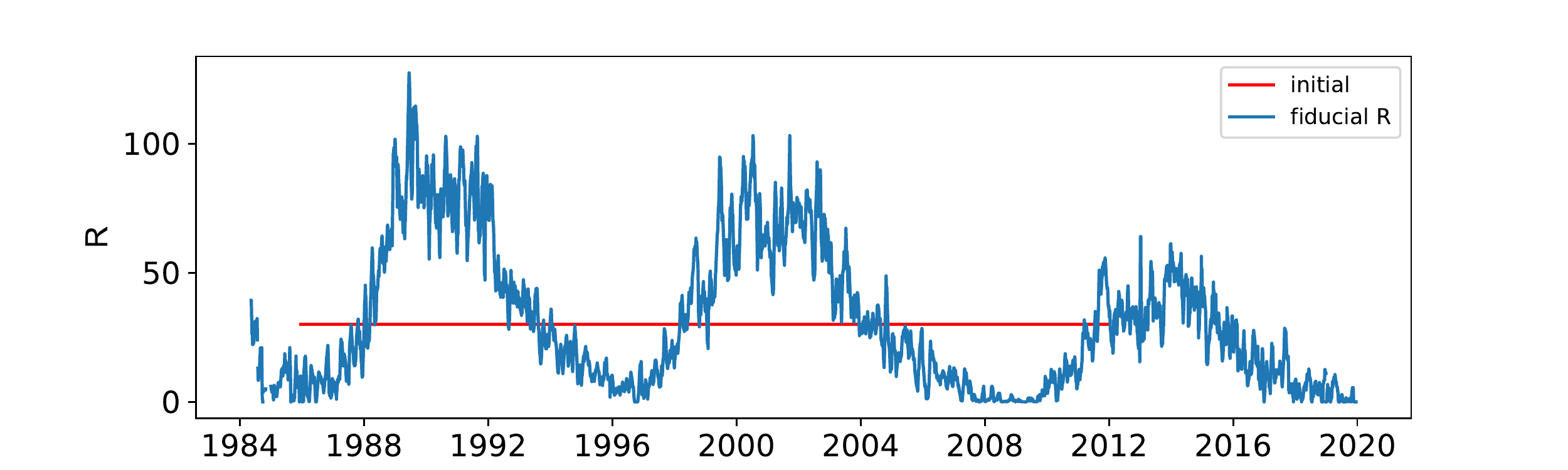}
    \caption{Comparisons of the sunspot number from our fiducial solution with the initial series coming from F.~Zloch (upper panel) and an arbitrary flat curve (bottom panel). }
    \label{fig:initial_comparison}
\end{figure}

The resulting target series $R_0$ consequently differs from the initial one considerably. This difference is plotted for instance in Fig.~\ref{fig:initial_comparison}, upper panel, where we took Franti\v{s}ek Zloch as an initial. One can see that even in the first evaluation windows between 1986--1999 the fiducial solution differs from the initial series. Due to the use of the observations of other observers, the target series is longer -- it starts in the years before the initial series was available and naturally extends beyond the end of the initial series. 

As a last note we would like to add that the sensitivity to the initial series is so low, that even with a flat initial series (that is a series where both $g$ and $f$ were constant) converged to the solution which had the same character as the others (see Fig.~\ref{fig:initial_comparison}, bottom panel). Note a considerably different amplitude as compared to the solution initiating from F.~Zloch. 

%\subsection{Sensitivity to the choice of window sizes}
%Describe the difference in the results when different choices of $W$ and $\Delta W$ are made. 
%
%NOTE: Table is possible, however given the amount of tables and figures already, I don't think it is necessary and only a wordy comment might suffice. 

% Tables:
% * correlation coefficients and slopes for a few choices as compared to the fiducial solution (5000/100).

%Figures:

% * show sampling effect in the conversion coefficients for a few observers

% HERE

\section{Concluding remarks}

In this paper we describe a new iterative methodology that allows to compute a target series of sunspot numbers. It does not rely on a single reference observer, it in fact uses as much information as possible. The methodology is very robust and except for the overall amplitude, it provides us with essentially the same solution regardless the initial series. 

Our methodology consists of modification of the current WDC-SILSO methodology. The main differences are the following:

\begin{itemize}
    \item We do not use a pilot station. Our resulting composite does not prefer any of the observers and constructs a series of the ``hypothetical'' stable observer. As a consequence, the overall amplitude is not constrained. When comparing with other concurrent series, it should be normalised, using a methodology similar to normalisation of one observer to other. 
    \item We split the determination of the personal scaling coefficients and the computation of the composite. Those two steps use the independent set of observations. 
    \item We use all the available observations and do not reject any of them on the statistical basis. This makes our methodology suitable for smaller networks with a rather scarce coverage. In a sense, it constitutes a compromise between the backbone method and the modern statistically driven methods. 
    \item We use two personal scaling coefficients of the observers, one for the number of groups and one for the number of sunspots. These coefficients are allowed to slowly change in time independently. 
    \item Our methodology is iterative and builds the target composite step-by-step. It always uses all the input data and naturally allows an inclusion of the newly discovered archival observations, should any appear later. 
\end{itemize}

\begin{figure}
    \centering
    \includegraphics[width=\textwidth]{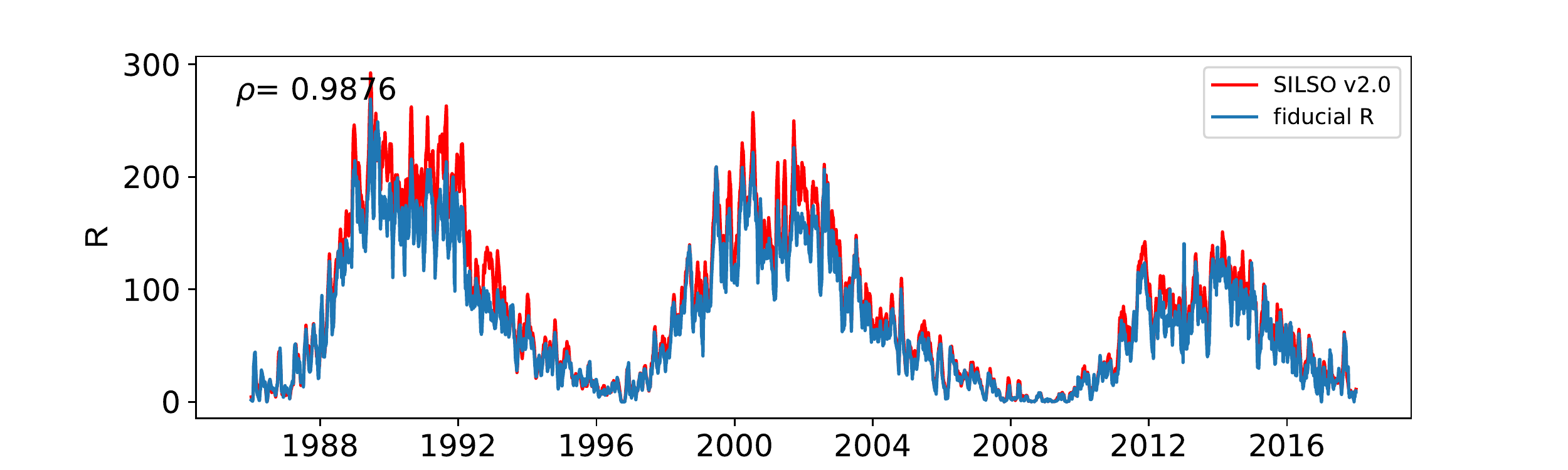}
    \caption{Comparions of the sunspot number from our fiducial solution with SILSO v2.0 sunspot number. }
    \label{fig:SILSO_comparison}
\end{figure}

\begin{figure}
    \centering
    \includegraphics[width=\textwidth]{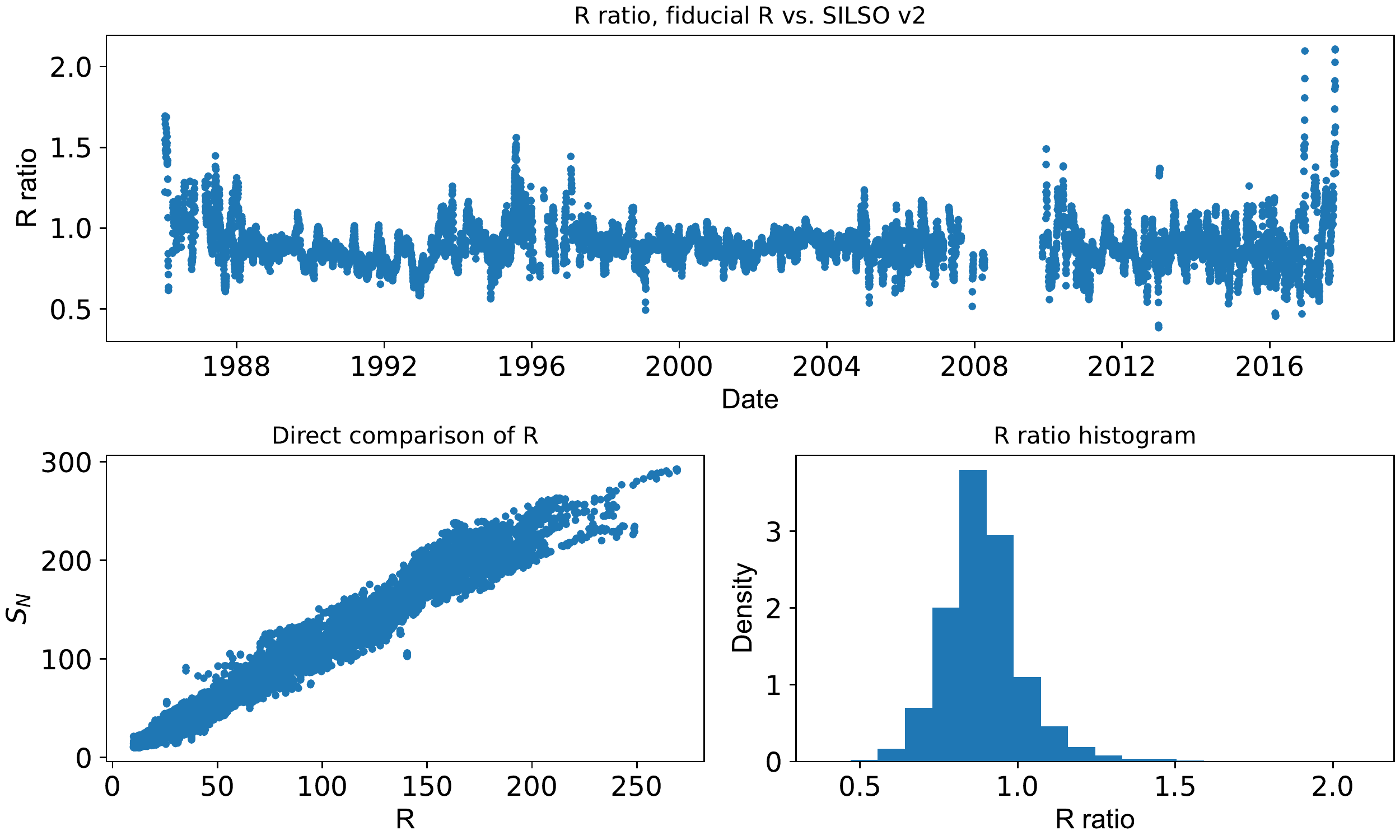}
    \caption{Similar plot to Fig.~\ref{fig:initials_Zloch_Schmied} directly comparing the values of our fiducial solution to SILSO v2.0 sunspot number.}
    \label{fig:SILSO_comparison_ratios}
\end{figure}

The question may be how does our computed relative number compare to the ``official'' international products. This can easily be demonstrated by plots and statistical properties. In Fig.~\ref{fig:SILSO_comparison} we show a comparison of our fiducial solution with F.~Zloch as initial and the sunspot number SILSO v2.0 \citep{2014SSRv..186...35C}. It can be seen that the overall agreement is very good, the Pearson's correlation coefficients reaches $\rho=0.99$. It is clear from the plot that our solution has a smaller overall magnitude, the scaling coefficient is approximately $0.863$. The ratio between the fiducial $R$ and SILSO v2.0 sunspot number is quite stable with time except for a few outliers in the minimum phase, the histogram of ratios shows the unimodality (Fig.~\ref{fig:SILSO_comparison_ratios}). 

\begin{figure}
    \centering
    \includegraphics[width=\textwidth]{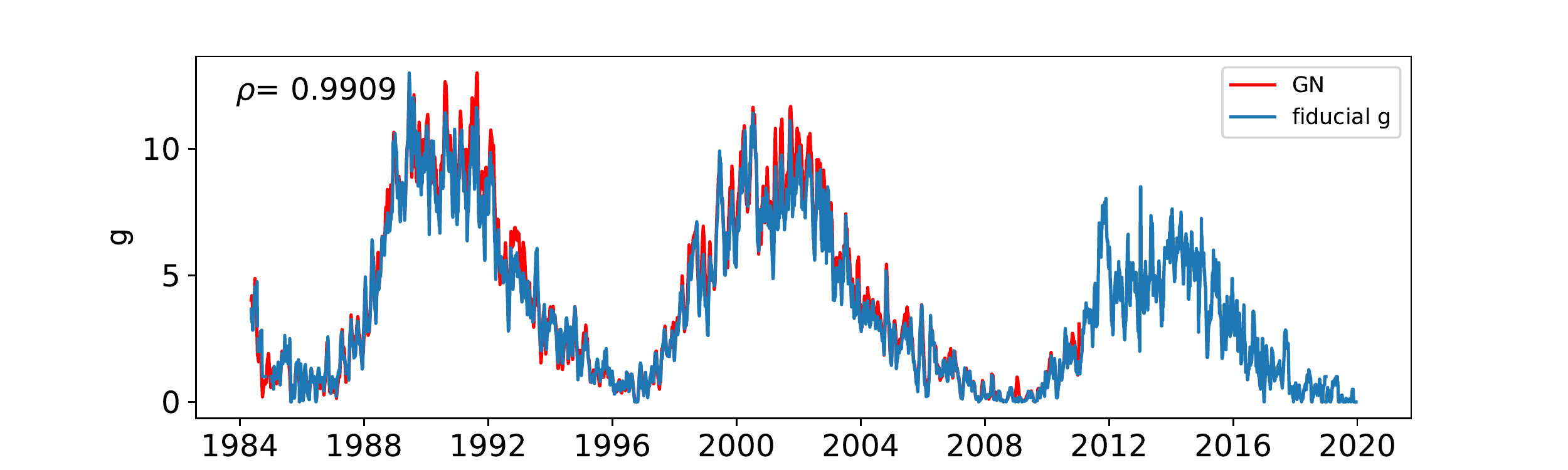}
    \caption{Comparisons of the group number from our fiducial solution with revised group number. (Note that revised group number is only available until 2010, the correlation coefficient is computed for the overlapping period.) }
    \label{fig:GN_comparison}
\end{figure}

\begin{figure}
    \centering
    \includegraphics[width=\textwidth]{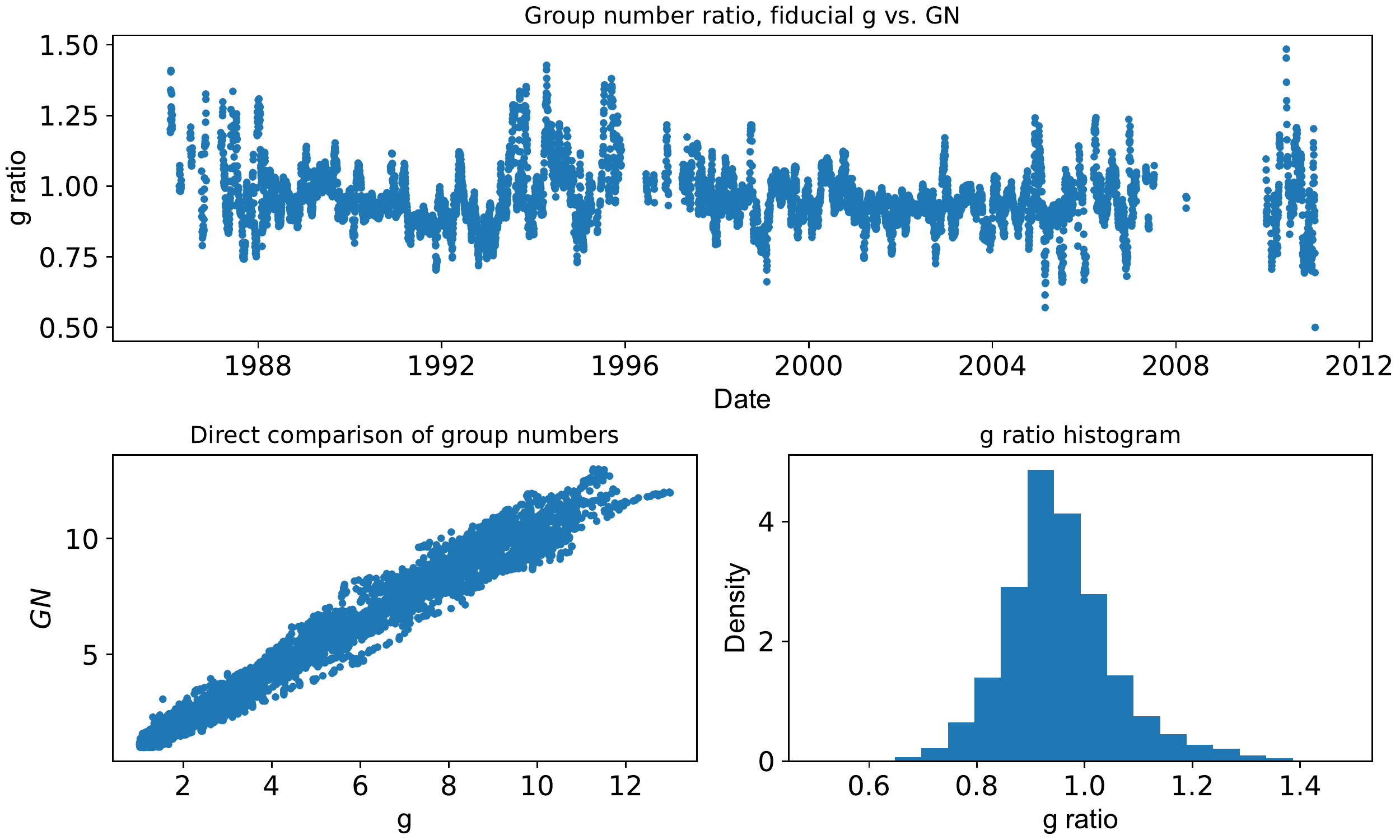}
    \caption{Similar plot to Fig.~\ref{fig:initials_Zloch_Schmied} directly comparing the values of our fiducial solution group number to \cite{2016SoPh..291.3061V}'s group number.}
    \label{fig:GN_comparison_ratios}
\end{figure}

SILSO sunspot number is not the only one available in the community. A solar index of a slightly different nature is the group number GN, introduced by \cite{1998SoPh..179..189H}. We compare this index recalculated by \cite{2016SoPh..291.3061V} with a number of groups, which is as a by-product computed also by our method. Fig.~\ref{fig:GN_comparison} shows an excellent agreement between the two series, again, the ratio is stable over the period for available for comparison (Fig.~\ref{fig:GN_comparison_ratios}).  

The code returns not only the resulting reference $g_0$, $f_0$ and $R_0$, however it naturally keeps and returns the personal conversion coefficients of each observer. It therefore allows to study the performance of individual observers \emph{ex-post} and allows to control possible overall trends that would invalidate the solution. 

\begin{figure}
    \centering
    \includegraphics[width=0.49\textwidth]{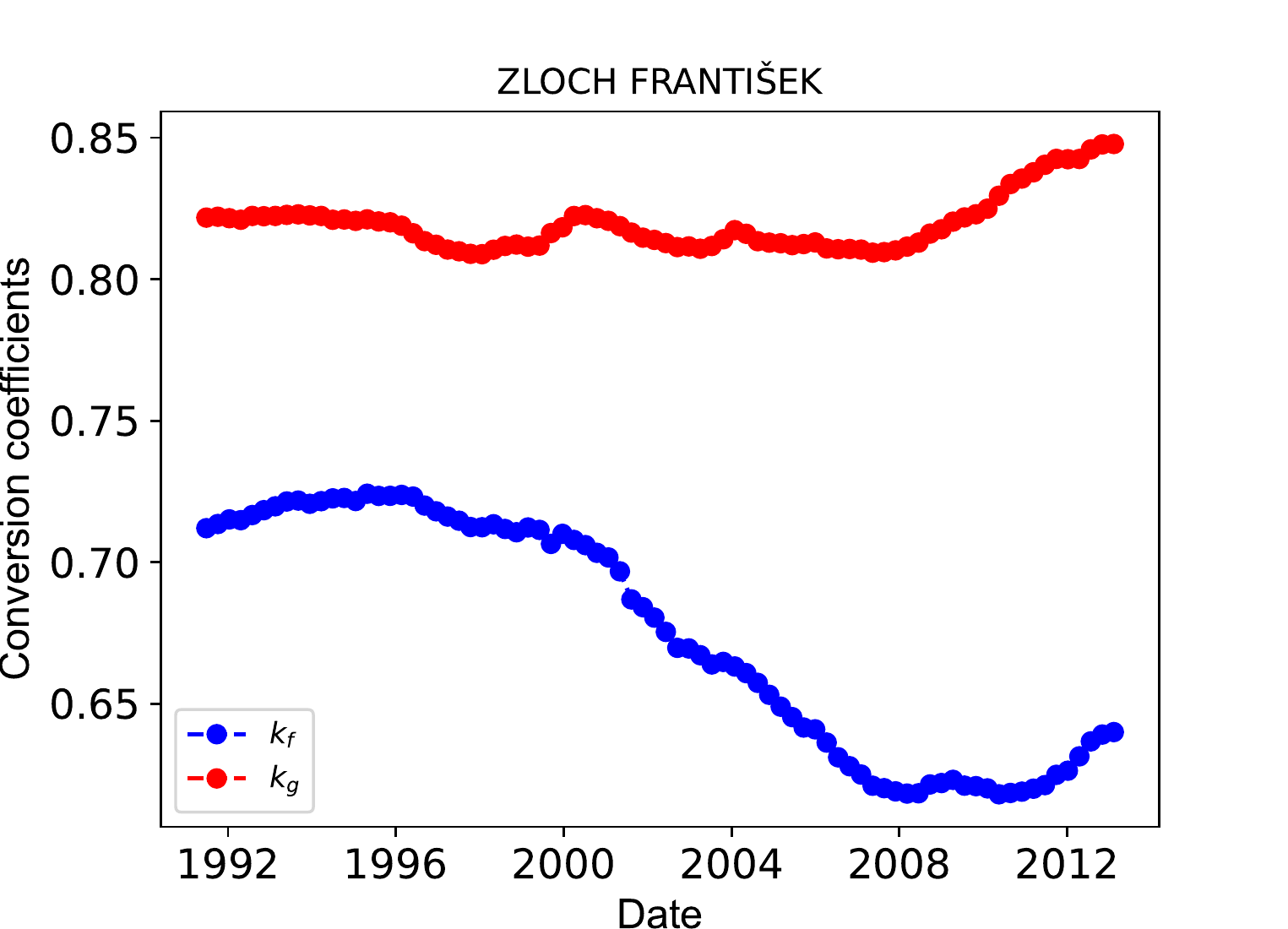}
    \includegraphics[width=0.49\textwidth]{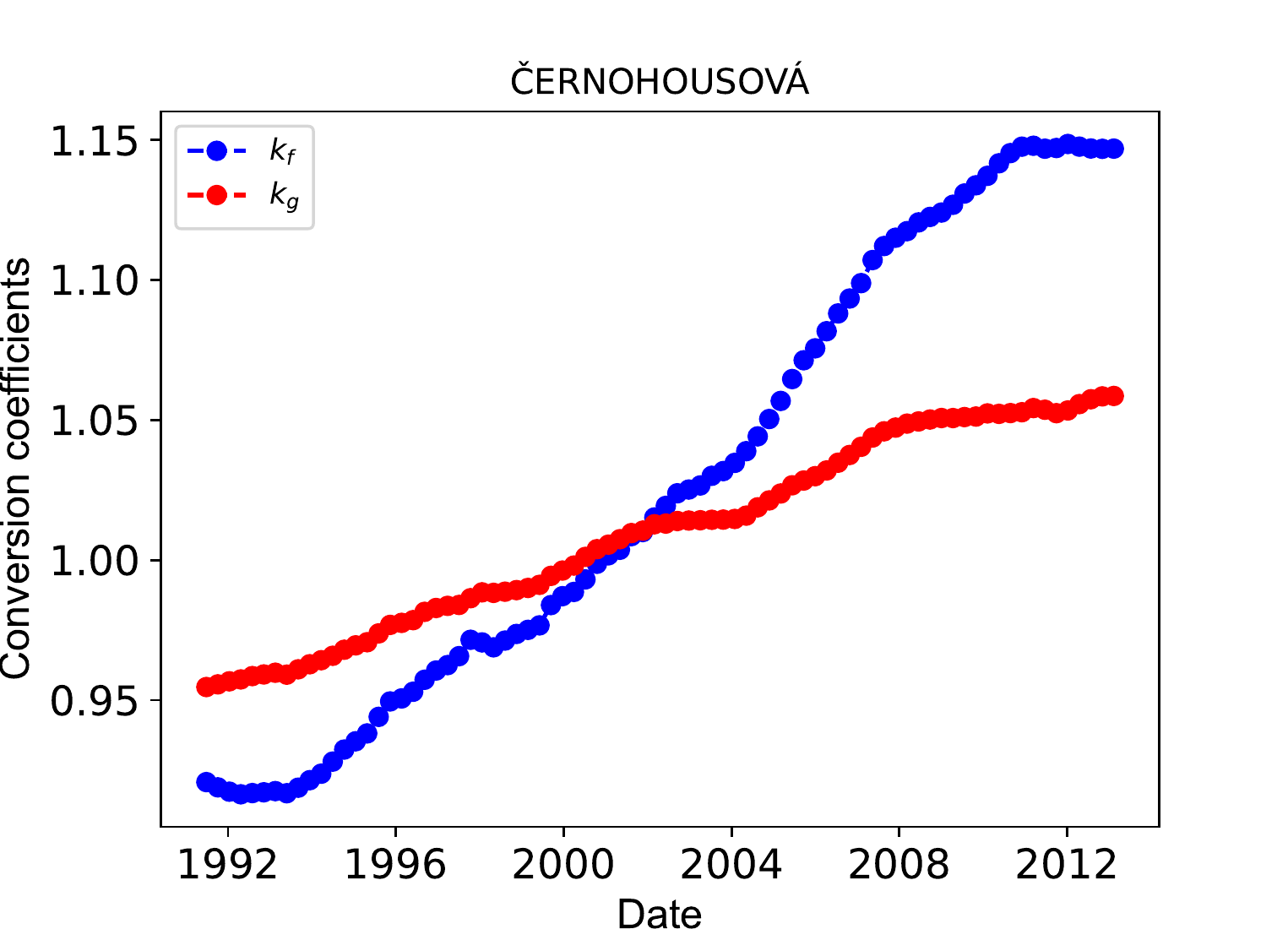}
    \caption{Example evolutions of conversion coefficients for selected long-term observers. }
    \label{fig:example_ks}
\end{figure}

In Fig.~\ref{fig:example_ks} we plot personal conversion coefficients of two observers used in our study. They both depict a gradual change. We need to note here that the apparently smooth curve is over-sampled due to the selection of our fiducial solution ($W=5000$~days, $\Delta W=100$~days). The effective time resolution is $W/2=2500$~days, which is about 6.8~years. Either way, a decrease of $k_f$ after year about 2000 is evident in the left panel. It would mean that F.~Zloch suddenly systematically saw considerably larger number of sunspots, whereas his $k_g$ remained more-or-less the same, that is, he at the same time registered a constant number of sunspot groups. 

In fact, it is possible to explain this gradual change in $k_f$. It has become a practice at Solar patrol at the end of 1990s to use an \emph{ex-post} camera aid to cross-check the drawing obtained by projection using a small 63-mm telescope with the direct observations by 20-cm telescope in a large magnification. This hypothesis is confirmed by the fact that the personal conversion coefficients of the other Solar patrol observers active at the same time showed similar trends. This change in observing methodology was not written in the logs, either way it became a common practice that was used until this generation of observers retired. The new observers at Solar patron do not follow this practice and rely only on projection by the 63-mm telescope. 

The other displayed trends of the observer Bo\v{z}ena \v{C}ernohousov\'a from Prost\v{e}jov observatory indicated constantly growing personal conversion coefficients, $k_f$ with a larger growth rate than $k_g$, which very likely indicates a gradually worsening vision. 

\begin{figure}
    \centering
    \includegraphics[width=\textwidth]{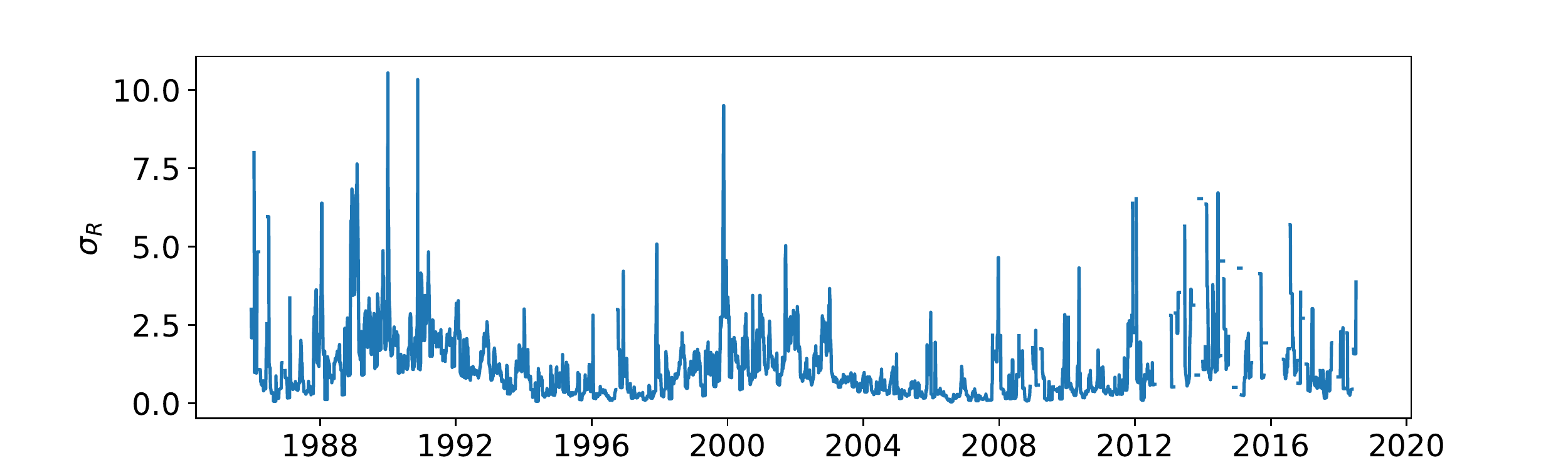}
    \caption{Derived statistical uncertainty of the fiducial $R$. }
    \label{fig:sigmaR}
\end{figure}

Lastly, we would like to point out that since the reference values of $g_0$ and $f_0$ are obtained by means of sample averaging, it is possible to compute formal statistical uncertainties of those two variables. By following the principle of propagation of uncertainties \citep{enwiki:1088872058} we hence obtain statistical uncertainty of $R_0$ as well. For our fiducial solution we plot the uncertainties of $R_0$ in Fig.~\ref{fig:sigmaR}. 

\begin{acks}
We would like to dedicate this paper to Franti\v{s}ek Zloch (29.\,1.\,1949\,--\,11.\,2.\,2022), a meticulous solar observer at the Solar patrol of the Astronomical Institute of the Czech Academy of Sciences. He has held the project FOTOSFEREX for a very long time in times when the access to the international observations was close to impossible and a need for the local network of solar observers was eminent. We would also like to thank the anonymous referee, who gave us many comments improving the paper. 
\end{acks}

\section{Additional statements}

\begin{authorcontribution}
M\v{S} designed the research, wrote the code, processed the data, wrote an initial manuscript draft. MP and BS digitised the observations. JD contributed the methodology from the mathematical point of view. All authors read and contributed to all manuscript versions. 
\end{authorcontribution}

\begin{fundinginformation}
Authors from the Astronomical Institute of the Czech Academy of Sciences were supported by the institution project ASU:67985815.
\end{fundinginformation}

\begin{dataavailability}
The data are available upon request at the corresponding author. The compiled database used in this study is available as an example via GitHub repository. \\
\url{https://github.com/michalsvanda/sunspot\_numbers}. 
\end{dataavailability}

%\begin{materialsavailability}
%Information about available material ...
%\end{materialsavailability}

\begin{codeavailability}
The code \citep{THECODE} is available via the GitHub repository.\\ \url{https://github.com/michalsvanda/sunspot\_numbers}
\end{codeavailability}

\begin{ethics}
\begin{conflict}
The authors declare that they have no conflicts of interest.
\end{conflict}
\end{ethics}

\bibliographystyle{spr-mp-sola}
\bibliography{biblio}

\end{article}

\end{document}